\documentclass[useAMS]{mn2e}

\usepackage{amssymb,amsmath,psfig,times}
\def\gsim{ \lower .75ex \hbox{$\sim$} \llap{\raise .27ex \hbox{$>$}} }
\def\lsim{ \lower .75ex\hbox{$\sim$} \llap{\raise .27ex \hbox{$<$}} }

\def\sc{Schwarzschild}
\def\beq{\begin{equation}}
\def\eeq{\end{equation}}

\def\sc{Schwarzschild}

\newcommand{\etal}{{et al.~}}

\title[CMB quenching of high--redshift radio AGNs]
{CMB quenching of high--redshift radio--loud AGNs}
\author[G. Ghisellini et al.]
{G. Ghisellini$^1$\thanks{Email: gabriele.ghisellini@brera.inaf.it}, 
F. Haardt$^{2,3}$, B. Ciardi$^4$, T. Sbarrato$^{5}$, E. Gallo$^{6}$, F. Tavecchio$^{1}$, A. Celotti$^{7, 1, 8}$
\\
$^1$ INAF -- Osservatorio Astronomico di Brera, Via Bianchi 46, I--23807 Merate, Italy\\
$^2$ DiSAT, Universit\`a dell'Insubria, via Valleggio 11, I--22100 Como, Italy \\
$^3$ INFN, Sezione di Milano--Bicocca, Piazza della Scienza 3, I--20126 Milano, Italy \\
$^4$ Max Planck Institute for Astrophysics, Karl--Schwarzschild Strasse 1, D--85741 Garching, Germany \\
$^5$ Dipartimento di Fisica ``G. Occhialini", Universit\`a di Milano--Bicocca, P.za della Scienza 3, I--20126 Milano, Italy\\ 
$^6$ Department of Astronomy, University of Michigan, 500 Church St., Ann Arbor, MI 48109, USA \\
$^7$ Scuola Internazionale Superiore di Studi Avanzati, Via Bonomea 265, I--34135 Trieste, Italy \\
$^8$  INFN--Sezione di Trieste, via Valerio 2, I-34127 Trieste, Italy \\
}

\begin{document}  

\maketitle

\begin{abstract}
The very existence of more than a
dozen of high--redshift ($z\gsim 4$) blazars indicates that a much larger 
population of misaligned powerful jetted AGN was already in place when the Universe 
was $\lsim$1.5 Gyr old. 
Such parent population proved to be very elusive, and escaped direct detection in radio surveys so far. 
High redshift blazars themselves seem to be failing in producing extended radio--lobes, raising questions 
about the connection between such class and the vaster population of radio--galaxies. 
We show that the interaction of the jet electrons with the intense cosmic microwave background (CMB) radiation 
explains the lack of extended radio emission in high redshift blazars and in their parent population, 
helping to explain 
the apparently missing misaligned counterparts of high redshift blazars. 
On the other hand, the emission from the more compact and more magnetised hot spots
are less affected by the enhanced CMB energy density.
By modelling the spectral energy distribution of blazar lobes and  
hot spots
we find that most of them should be detectable by low frequency 
deep radio observations, e.g., by LOw--Frequency ARray for radio astronomy (LOFAR) 
and by relatively deep X--ray observations with good angular resolution, e.g., by the {\it Chandra} satellite. 
At high redshifts, the emission of a misaligned relativistic jet, being de--beamed, is missed by current
large sky area surveys. 
The isotropic flux produced in the hot spots can be below  $\sim 1$ mJy
and the isotropic lobe radio emission is quenched by the CMB cooling.  
Consequently, even sources with very powerful jets can go undetected in 
current radio surveys, and misclassified as radio--quiet AGNs.
\end{abstract}

\section{Introduction}

Relativistic jets from powerful radio--loud Active Galactic Nuclei (AGN) carry energy and particles 
from within the sphere of influence of the central super--massive black hole out to distances that far 
exceed the size of their host galaxy (e.g. Schoenmakers et al. 2000; de Vries, Becker \& White 2006). 
When the jet interacts with the external medium, a hot spot is formed, powering
the extended structures we call ``lobes".
The lobes are characterized by relatively low magnetic fields (tens of $\mu$G), and are 
thus inefficient synchrotron radiators. 
Minimum energy arguments, based on the assumption of equipartition between the 
particle and magnetic energies, suggest that the lobes  
of the most powerful sources can store up to $10^{61}$ erg in energy.  

Even when we do not have (yet) resolved radio maps of the
hot spots and of the lobes (e.g. when we have only single dish observations), 
their presence can be inferred from steep low frequency radio spectra, characteristic of optically 
thin synchrotron emission, in contrast to the flat spectra of partially self--absorbed, 
optically thick jets

\begin{figure} 
\vskip -0.6 cm
\psfig{file=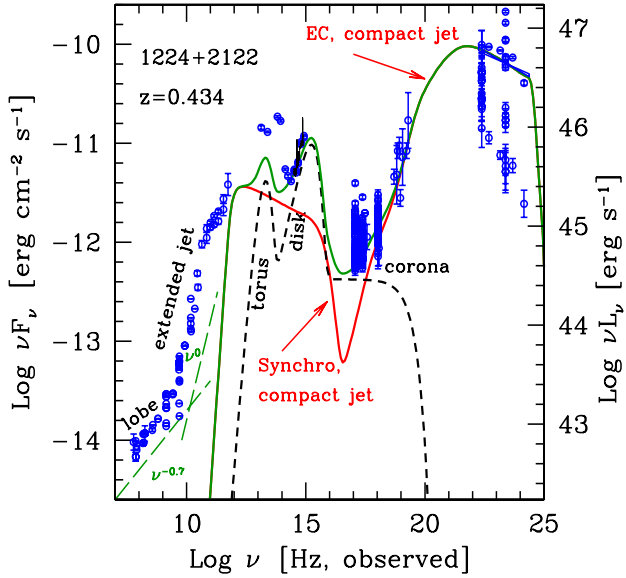,width=8.5cm,height=9cm } 
\vskip -0.5 cm
\caption{
SED of CRATES J1224+2122 (alias PKS 1222+21 alias 4C +21.35), a {\it Fermi}/LAT detected blazar 
(Shaw et al. 2012). 
The solid red line is the emission from the single--zone jet model we adopted 
(see text and Ghisellini \& Tavecchio 2015), 
the short-dashed black line is the contribution from the thermal components 
(i.e., accretion disc, the torus and the X--ray corona), while the solid green line is the sum of the two. 
The adopted jet model assumes that most of the emission comes from a compact jet region, and therefore
the synchrotron emission is self--absorbed up to $\sim 10^3$ GHz. 
At lower frequencies the flux is produced by several larger jet regions, self--absorbing at smaller
frequencies, and by the extended radio--lobes, whose synchrotron emission is optically thin.
The latter component is relatively steep ($F_\nu\propto \nu^{-0.7}$), while the partially opaque emission
from the jet is flat  ($F_\nu\propto \nu^0$), as illustrated by the long--dashed green lines. 
}
\label{1224}
\end{figure}

Fig. \ref{1224} illustrates the point by showing the broad band spectral energy distribution 
(SED) of 1224+2122 (a.k.a. 4C +21.35), a $z$=0.434 blazar detected in the $\gamma$--ray band 
by the {\it Fermi} satellite (Ackermann et al. 2011; Shaw et al. 2012)
\footnote{All shown data points are taken from the ASI Science Data Center (ASDC) archive.}.
Multiple data points at the same frequency indicate the strong variability typical of blazars, 
that can reach an amplitude of 1--2 orders of magnitude at X--ray and $\gamma$--ray energies, while 
it is much less extreme in the radio band. 
The signature of 
isotropic and optically thin synchrotron emission produced by the hot spots and by the 
radio lobes is the relatively steep slope of the radio spectrum 
below a few GHz, $F_\nu \propto \nu^{-0.7}$-- $\nu^{-1.5}$.
Above a few GHz the radio spectrum, produced by the
partially opaque jet (labelled as ``extended jet" in Fig. \ref{1224}), 
is instead much flatter, i.e., $F_\nu \propto \nu^0$. 
Note that the long--dashed lines at the lowest radio frequencies are not fits to the 
lobe and extended jet emission. 
Their purpose here is simply to guide the eye.
Note that the case of 1224+2122 is not common, since most blazars
do not show the spectral steepening at low radio frequencies, even
when going down to 74 MHz, as shown by Massaro et al. (2013).

Whereas the core jet emission is strongly collimated and boosted by relativistic beaming,
the hot spot and lobe emission is largely isotropic (or at most mildly beamed). 
This implies that, as a first approximation, the jet--to--lobe flux ratio (i.e., the 
so--called ``core dominance") is an indicator of the observer's viewing angle. 
For each source whose jet is pointing at an angle $\theta_{\rm v}\le 1/\Gamma$ (where $\Gamma$ 
is the jet bulk Lorentz factor), and whose power is large enough to produce radio lobes, there 
must exist $\sim 2\Gamma^2$ sources with strongly misaligned (and thus undetectable) jets, and whose radio 
lobes should be visible regardless of the viewing angle, given the required instrument sensitivity. 
If so, then radio--galaxies and misaligned radio--loud quasars ought to be $2\Gamma^2$ times more numerous than blazars.   

Volonteri et al. (2011) showed that this straightforward prediction is confirmed up to $z\sim 3$, 
beyond which the number of radio--galaxies dramatically drops. 
More specifically, the expected number density of misaligned radio--loud quasars and radio--galaxies 
as inferred from the {\it Swift}/BAT sample of luminous, massive blazars (Ajello \etal 2009),
overestimates the number of observed luminous, radio--loud AGNs (from SDSS--DR7+FIRST)  
by a factor $\sim 3$ in the redshift bin $z=3-4$, and by a factor $>$10 between $z=4-5$ 
(see Table 1 and Fig. 3 in Volonteri \etal 2011; qualitatively similar conclusions are 
reached by Kratzer \& Richards 2015; see also Haiman, Quataert \& Bower 2004 and Mc Greer Helfand \& White  2009). 

Volonteri et al. (2011) put forward some possibile explanations of this discrepancy, namely:  
(i) heavy optical obscuration of high--$z$ radio galaxies; 
(ii) substantial drop in the average $\Gamma$ for high--$z$ sources; and  
(iii) substantial dimming of the radio lobes at $z\gsim 3$. 
The first scenario implies the existence of a large population of infrared--luminous, 
radio--loud quasars with weak optical counterparts.
The second scenario implies a very large density of the emitting
relativistic electrons (leading to a very large jet kinetic power; 
see Ghisellini \& Tavecchio 2010). 

With respect to the third hypothesis, the idea has long been entertained of 
Cosmic Microwave Background (CMB) photons affecting the behavior of jetted AGNs (e.g. Celotti \& Fabian 2004). 

In a recent work, Ghisellini et al. (2014a) have explored specifically how the interaction between 
the CMB radiation and electrons within the jet--powered lobes affects the appearance of 
jetted AGNs at different redshifts. 
The major results can be summarized as follows: owing to its $(1+z)^4$ dependence, 
the CMB energy density $U_{\rm CMB}$ starts to dominate over the magnetic energy density $U_{\rm B}$ 
within the lobes above $z \simeq$3, thereby suppressing the synchrotron radio flux at higher $z$ 
(hereafter referred to as {\it CMB quenching}). 
At the same time, high--energy electrons will cool effectively by 
Inverse Compton losses scattering off CMB photons.
Combined, these two effects result in a significant enhancement of the diffuse 
X--ray emission -- in the form of X--ray lobes --  from high--$z$ quasars 
(Celotti \& Fabian 2004; Mocz, Fabian \& Blundell 2011).
The hot spot radio emission is less affected by the CMB quenching,
since they are more compact and more magnetised than the lobes.
However, at low redshifts, their contribution to the total
radio flux is less than the contribution due to the radio lobes.

In this paper, we consider all radio--loud sources at $z>4$ confirmed to be blazars,
and construct their broad band SEDs.
These SEDs have all a radio flat spectrum (where flat means
$F(\nu)\propto \nu^{-\alpha}$ with $\alpha<0.4$).
This spectrum is produced by different regions of the jet, from the very
compact (subparsec) to the extended regions (tens of kpc from the black hole, or more).
In all cases we see no hints of the presence of 
a steep thin synchrotron component produced by the hot spots and by the radio lobes
at (rest frame)  frequencies $\lsim$1 GHz, hence providing observational 
support to the CMB quenching scenario for high--$z$ jetted AGNs. 
By applying a state--of--the--art jet emission model, we will fully characterise the physical properties 
of the jets. 
This will allow us to constrain the expected properties of 
both the hot spots and the
lobes, whose broad band emission can then be modelled under few reasonable 
assumptions (such as energy equipartition between electrons and magnetic field). 
Finally we will address the observability of such low--surface brightness radio--lobes from the known 
population of $z>4$ blazars at (rest--frame) frequencies at or below the GHz band, specifically with 
the LOw Frequency ARray (LOFAR, van Haarlem et al., 2013).

We adopt a cosmology with $\Omega_{\rm m}=0.3$, $\Omega_\Lambda=0.7$, and 
$H_0=70$ km s$^{-1}$ Mpc$^{-1}$. 

\begin{table*} 
\centering
\begin{tabular}{l l c r r l l l l r l}
\hline
\hline
Name &$z$ &SDSS+FIRST &$R$ &$L_{\rm bol}$ &$L_{\rm CIV}$  &$M_{\rm vir}$  &$L_{\rm BLR, 45}$ &$M_{\rm fit}$ 
&$L_{\rm d, 45}$    &Ref\\
~[1] &[2] &[3] &[4] &[5] &[6] &[7] &[8] &[9] &[10] &[11] \\
\hline   
032444.30 --291821.0 &4.630 &N &4468  &...  &...   &...   &...  &5e9    &200   &Yuan et al. 2006 \\
052506.18 --334305.5 &4.413 &N &1230  &...  &...   &...   &...  &3e9    &148   &Fabian et al. 2001b\\
083946.22 +511202.8	 &4.390 &Y &285   &47.5 &45.0  &8.9e9 &8.8  &7e9    &178   &Bassett et al. 2004; Sbarrato et al. 2013a  \\
090630.75 +693030.8  &5.47  &N &1000  &...  &...   &...   &...  &3e9    &68    &Romani et al. 2004\\
102623.61 +254259.5  &5.304 &Y &5200  &...  &...   &...   &...  &5e9    &75    &Sbarrato et al. 2012 \\ 
102838.80 --084438.6 &4.276 &N &4073  &...  &...   &...   &...  &4e9    &120   &Yuan et al. 2000\\ 
114657.79 +403708.6  &5.005 &Y &1700  &...  &...   &...   &...  &4e9    &114   &Ghisellini et al. 2014b  \\
125359.62 --405930.5 &4.460 &N &4700  &...  &...   &...   &...  &2e9    &42    &Yuan et al. 2006\\
130940.70 +573309.9  &4.268 &Y &133   &47.1 &45.1  &2.1e8 &46.0 &8e9    &57    &This paper\\
132512.49 +112329.7  &4.412 &Y &879   &47.4 &45.4  &2.8e9 &46.4 &3e9    &55    &This paper; Sbarrato et al. 2013a\\
142048.01 +120545.9  &4.034 &Y &1904  &47.1 &45.0  &1.9e9 &8.16 &2e9    &54    &Sbarrato et al. 2015 \\ 
143023.7  ~~~+420436 &4.715 &Y &5865  &...  &...   &...   &...  &1.5e9  &135   &Fabian et al. 2001a\\  
151002.92 +570243.3  &4.309 &Y &13000 &47.1 &44.9  &3.2e8 &6.6  &1.5e9  &59    &Yuan et al. 2006   \\
171521.25 +214531.8  &4.011 &Y &30000 &...  &...   &...   &...  &6e8    &5.5   &Gobeille et al. 2014; Wu et al. 2013\\ 
213412.01 --041909.9 &4.346 &Y &24000 &...  &...   &...   &...  &1.5e9  &101   &Sbarrato et al. 2015 \\
222032.50 +002537.5  &4.205 &Y &4521  &46.9 &45.1  &1.4e9 &10.  &2e9    &45    &Sbarrato et al. 2015 \\
\hline
\hline
\end{tabular}
\caption{
Our blazar sample, containing all known blazars at $z\ge4$.
Col. [1]: right ascension and declination;
Col. [2]: redshift;
Col. [3]: flag for belonging to the SDSS+FIRST survey;
Col. [4]: radio--loudness;
Col. [5]: logarithm of the bolometric luminosity (from S11) 
Col. [6]: logarithm of the CIV emission line luminosity (from S11);
Col. [7]: black hole mass (in solar masses, best estimate from S11);
Col. [8]: Broad Line Region luminosity (in units of $10^{45}$ erg s$^{-1}$), obtained  from CIV;
Col. [9]: black hole mass (in solar masses) from disk fitting;
Col. [10]: accretion disk luminosity (in units of $10^{45}$ erg s$^{-1}$) from disk fitting;
Col. [11]: References. 
}
\label{sample}
\end{table*}

\section{The blazar sample}

Following Sbarrato et al. (2013a), we consider all 31 spectroscopically confirmed $z>4$ quasars 
in the SDSS+FIRST sample owning a radio--loudness $R=F_{5 {\rm GHz}}/F_{2500{\rm \AA}}>100$, 
where $F_{5 {\rm GHz}}$ and $F_{2500{\rm \AA}}$ are the monochromatic fluxes at 5 GHz and at 2500  
\AA, respectively (see Shen et al. 2011, thereafter S11). 
Six bona--fide blazars (see Table \ref{sample}) are then identified from this initial 
sample of 31 sources  based on {\it Swift} X--Ray Telescope (Sbarrato et 
al. 2012, 2015; Ghisellini et al. 2014b; 2015) and {\it NuSTAR} 
follow--up observations (Sbarrato et al. 2013b). 
Two other quasars out of the 31 had archival X--ray data, 
showing intense and hard X--ray spectra: by fitting their multiwavelength 
SEDs, we can now classify 
SDSS J130940.70+573309.9 and SDSS J132512.49+112329.7 
as blazars, thanks to their small viewing angles and high Lorentz factors. 

In addition, we include 8 other blazars at $z>4$ (see Table \ref{sample}) serendipitously 
identified in X--rays. 
Among these 8 sources we note that 213412.01--041909.9 is in the SDSS+FIRST sample, 
but it was not included in the Shen \etal sample because there was no spectroscopic follow--up. 
A spectroscopic redshift of $z=4.346$ was determined by Hook \etal (2002) for this source. 
The same occurs for 143023.7+420436 and SDSS J171521.25+214531.8, 
whose redshifts ($z$=4.715 and $z$=4.011, respectively)  were first 
determined by Hook \& McMahon (1998). 
The latter has a further indication of its orientation given by extended 
radio emission (Gobeille et al. 2014). 

To summarize, we consider a grand total of 16 blazars with confirmed  $z>4$; 3 of them are at 
$z>5$, and 11 objects belong to the SDSS+FIRST sample. 
In Tab. \ref{sample} we list the main references for each object, while the SEDs are shown in 
Fig. \ref{1715} and Fig. \ref{sed}\footnote{Data are taken from the papers listed in Tab. \ref{sample} 
and from the ASDC archive at {\tt http://tools.asdc.asi.it/.}}.


\section{Emission models}

In this section we summarize the main properties of the models for
the jet, the hot spot and lobe emissions we use.
Full details can be found in Ghisellini \& Tavecchio (2009, for the jet)
and in Ghisellini et al. (2014a, for the lobe), and interested readers are referred there. 

Modelling the jet emission allows us to estimate the total jet power, that we assume to be conserved 
along the jet and to power the lobes on a much larger scale. 
A clear estimate of the jet power is therefore instrumental in modelling the lobe emission. 
Other important parameters characterising the lobes, such as the physical extension, 
the average magnetic field, and the particle energy distribution are, basically, unknown. 
We assume physically motivated values for them, and discuss how our choices impact on the final results. 

\subsection{Jet emission}

For the jet emission we adopt a simple, one--zone, leptonic model (Ghisellini \& Tavecchio 2009). 
The model assumes that most of the observed radiation is produced in a single
spherical region within a conical jet of semi-aperture $\psi =0.1$ rad. 
The spherical region is initially located at a distance $R_{\rm diss}$ from the black hole, 
it is homogeneously filled with a tangled magnetic field $B$, and it moves with velocity $\beta c$, 
corresponding to a bulk Lorentz factor $\Gamma$, at an angle $\theta_{\rm v}$ with respect to the line of sight. 
The resulting Doppler factor is $\delta=1/[\Gamma(1-\beta\cos\theta_{\rm v})]$. 
Relativistic electrons are injected throughout the spherical region at a constant rate $Q(\gamma)$,  
with total {\it comoving} power (i.e., as measured in the jet comoving frame)
\begin{equation}
P^\prime_{\rm e} = V m_{\rm e} c^2 \int_{\gamma_{\rm min}}^{\gamma_{\rm max}} Q(\gamma)\gamma d\gamma,
\end{equation}
where $V$ is the volume of the emitting region and $\gamma_{\rm min}$ and $\gamma_{\rm max}$ are the
minimum and maximum injection energies. 
The electron energy distribution $Q(\gamma)$ is taken as a smoothly connected  double power law, with 
slopes $s_1$ and $s_2$ below and above the break energy $\gamma_{\rm b}$, respectively: 
\begin{equation}
Q(\gamma)  \, = \, Q_0\, { (\gamma/\gamma_{\rm b})^{-s_1} \over 1+
(\gamma/\gamma_{\rm b})^{-s_1+s_2} } \quad {\rm [cm^{-3} s^{-1}]}.
\label{qgamma}
\end{equation}
In the following, we will assume that $s_1\leq 1$, $s_2>2$ and $\gamma_{\rm min}=1$. 
For this choice of $s_2$, the exact value of $\gamma_{\rm max}$ is not critical, since 
it corresponds to the end of the steep tails of both the synchrotron and the inverse Compton spectrum.
Having specified the injection term, the particle density distribution $N(\gamma)$ [cm$^{-3}$]
is found by solving the continuity equation at a time equal to the light crossing time (that is also 
the time--scale for doubling the source size), 
taking into account radiative cooling and electron--positron pair production. 
Additionally, we consider the presence of a standard, optically thick, geometrically thin accretion disk 
(Shakura \& Sunyaev 1973), assumed to emit a total luminosity $L_{\rm d} =10^{45} L_{\rm d, 45}$ erg s$^{-1}$.
The assumption of a moving blob may correspond to the ``internal shock" scenario,
in which two shells, initially separated by some distance $\Delta R$,
move with different $\Gamma$ and collide at a distance $\approx \Gamma^2 \Delta R$
from the black hole (Rees 1978; Ghisellini 1999; Spada et al. 2001).
Calculating the particle distribution at the light crossing time, when the particle injection
stops, corresponds to the maximum produced flux, and allows to neglect
adiabatic losses, and changes in the size, magnetic field and particle density of the blob. 

We further assume that a broad line region (BLR), responsible for reprocessing 10\% of $L_{\rm d}$, 
lies at a distance $R_{\rm BLR} = 10^{17}L^{1/2}_{\rm d, 45}$ cm. 
Further out, at $R=2.5\times 10^{18} L^{1/2}_{\rm d, 45}$ cm, a molecular torus intercepts 
and re--emits a fraction ($\sim$20--40\%) of $L_{\rm d}$ in the infrared band.

We consider the following emission processes within the jet: 
(i) synchrotron; (ii) synchrotron self--Compton (SSC);  and
(iii) inverse Compton of the relativistic electrons scattering off photons produced
by the disc, the BLR, and the dusty torus (collectively dubbed EC, for External Compton).
Photon--photon interactions and pair production are also accounted for, albeit these
processes turn out to be unimportant for our blazar sample, which is uniformly characterised 
by a steep $\gamma$--ray spectrum.

The source size $\psi R_{\rm diss}$ is constrained by:
(i) the minimum variability time--scale $t_{\rm var}$, as the size must be smaller than 
$\sim c\,t_{\rm var}\delta/(1+z)$; 
(ii) the inverse Compton to synchrotron luminosity ratio, which is a
function of $R_{\rm diss}$ (see Ghisellini \& Tavecchio 2009); and
(iii) the location of the peak frequency of the Compton component,
which, in turn, dictates the nature of the primary seed photons (i.e., IR from the torus vs. 
UV from the broad lines).

The resulting source size is too compact to account for the observed radio
emission, since synchrotron radiation is self--absorbed up to (observed) frequencies $\sim 10^3$ GHz. 
At lower frequencies, the observed flat radio spectrum 
($F_\nu \propto \nu^{-\alpha}$, with $\alpha<0.4$, and often $\alpha\sim 0$) 
is understood to be produced by the superposition of the emission from several, 
larger, parts of the jet. 
We do not aim to model this part of the spectrum, and 
in all SEDs shown in Figs. 1 through 6 the long--dashed lines corresponding to 
$F_\nu \propto \nu^0$ or to $F_\nu \propto \nu^{-1/3}$ simply guide the eye. 
The outer jet, responsible for the radio spectrum, can be safely neglected
because the corresponding luminosity is only a small fraction of the total, and 
therefore we can assume that the jet retains the bulk of its power up to the lobe site.
The magnetic field of the extended part of the jet, up to the kpc size,
should be large enough to be unaffected by the CMB quenching: assuming a jet
at $z\sim 4$ with $B\sim$1 G at a distance $d\sim 10^{18}$ cm and $B\propto d^{-1}$,
we have that the the jet magnetic field starts to be smaller than $B_{\rm CMB}$ 
for $d>0.4 /(\Gamma/10)$ kpc, where $B_{\rm CMB}$ is the magnetic field
in equipartition with the CMB (as seen in the comoving frame of the jet)
given in Eq. \ref{bcmb}.
If we further assume that the self absorption frequency of the different jet components
scales as $d^{-1}$, we expect that CMB will reduce the synchrotron flux produced by the extended
part of the jets for $\nu\lsim$ 250 MHz. 
This crude estimate shows that the jet emission remains unaltered at all but the very small
frequencies.

The jet carries power in various forms, all of which can be conveniently expressed as energy fluxes as
\begin{equation}
P_{\rm i} \, =\, \pi R^2 \Gamma^2\beta c\, U^\prime_{\rm i}.
\end{equation}
Here, $U'_{\rm i}$ represents the energy density (in the jet comoving frame) in radiation,
$U^\prime_{\rm r}$, corresponding to the radiative jet power $P_{\rm r}$, and in magnetic field, 
$U^\prime_{\rm B}=B^{\prime \, 2}/(8\pi)$, corresponding to the jet Poynting flux $P_{\rm B}$.  
Finally, we need to consider the kinetic energy density of the electrons, 
$U^\prime_{\rm e} = m_{\rm e}c^2\, \int N(\gamma)\gamma d\gamma$, corresponding to the power carried by the electrons, 
$P_{\rm e}$, and protons, $P_{\rm p}$ (we are assuming to have one cold proton -- i.e. whose kinetic energy is due to 
bulk motion only -- per emitting electron). 
The total jet power, $P_{\rm jet}$, is the sum of all these components.

\begin{table} 
\centering
\begin{tabular}{lllll lllll llll }
\hline 
\hline
$R_{\rm lobe}$ &$P_{\rm e,lobe,45}$ &$B_{\mu \rm G}$  &$\gamma_{\rm b}$ &$\log E_{\rm B}$ &$\log E_{\rm e}$ &Fig. \\
~[1] &[2] &[3] &[4] &[5] &[6] &[7]    \\
\hline   
50  &27  &24.5 &1e3  &59.56 &59.57   &\ref{1028r}, \ref{1028power}, \ref{1028gamma} \\
25  &27  &40.1 &1e3  &59.10 &59.10   &\ref{1028r}   \\
100 &27  &17.0 &1e3  &60.15 &60.14   &\ref{1028r}   \\
50  &82  &42.5 &1e3  &60.04 &60.04   &\ref{1028power} \\
50  &9   &14.1 &1e3  &59.08 &59.09   &\ref{1028power}  \\
50  &2.7 &7.8  &1e3  &58.57 &58.57   &\ref{1028power} \\
2  &27  &21.2 &1e4  &59.44 &59.43   &\ref{1028gamma}  \\
2  &27  &27.4 &1e2  &59.66 &59.66   &\ref{1028gamma}  \\
%
\hline
$R_{\rm HS}$ &$P_{\rm e,HS,45}$ &$B_{\mu \rm G}$  &$\gamma_{\rm b}$ &$\log E_{\rm B}$ &$\log E_{\rm e}$ &Fig. \\
~[1] &[2] &[3] &[4] &[5] &[6] &[7]    \\
\hline   
2  &27  &310 &1e3 &57.58 &57.57 &\ref{1028r}, \ref{1028power}, \ref{1028gamma} \\
1  &27  &585 &1e3 &57.22 &57.23 &\ref{1028r} \\ 
4  &27  &170 &1e3 &57.96 &57.95 &\ref{1028r} \\
2  &90  &540 &1e3 &58.06 &58.06 &\ref{1028power}  \\
2  &9   &180 &1e3 &57.11 &56.11 &\ref{1028power}\\
2  &2.7 &97  &1e3 &56.57 &56.57 &\ref{1028power}\\
2  &27  &95  &1e4 &56.55 &56.55 &\ref{1028gamma}  \\
2  &27  &98  &1e2 &56.58 &56.57 &\ref{1028gamma} \\
\hline
\hline 
\end{tabular}
\vskip 0.4 true cm
\caption{
Model parameters for the lobe (top part) and host spot (bottom part)
emission of 1028--0844 ($z$=4.276). 
For all models we assumed $s_1=-1$, $s_2=2.7$ and $\gamma_{\rm max}=10^6$.
Col. [1]: dimension of the emitting zone in kpc;
Col. [2]: power injected in each hot spot and lobe in the form of relativistic electrons in units of $10^{45}$ erg s$^{-1}$;
Col. [3]: magnetic field in $\mu$Gauss;
Col. [4]: break Lorentz factors of the injected electron distribution;
Col. [5] and [6]: logarithm of the magnetic and electron energy contained in the hot spot or in the 
lobe in units of erg;
Col. [7]: figure where the model is shown. 
}
\label{para4c}
\end{table}

\subsection{Hot spot emission}

Powerful jets deposit their kinetic energy into the intergalactic medium
through a termination shock, the hot spot, which in turn injects the power
received by the jet into the extended lobe structure.
To find out the main properties of the hot spot emission we will examine
first the hot spot properties of the closest powerful Fanaroff--Ryley II 
radio--galaxy, namely Cygnus A.
Then we will interpret the small, but resolved, radio emission of the distant blazar
SDSS J171521.25+214531.8 (1715+2145 hereafter) as due to the hot spot emission,
to check that if the properties of nearby and distant hot spots are similar for 
similar jet power.

\vskip 0.2 cm

\begin{figure}
\vskip -0.6 cm
\psfig{file=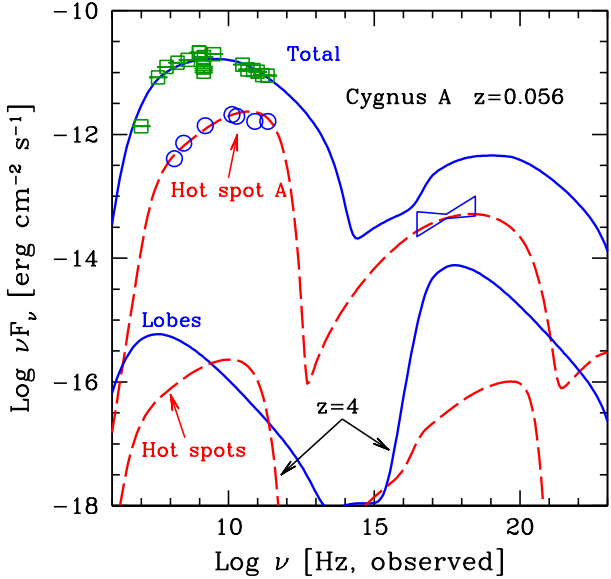,width=9cm,height=9.cm }
\vskip -0.8 cm
\caption{
SED of the hot spot A (the brightest) of Cygnus A (blue circles), 
compared to the total radio emission (green squares).
The blue bow tie refers to the X--ray flux of the hot spot A as observed by Chandra
(Wilson, Young \& Shopbell 2000).
The dashed red and solid blue lines are the corresponding SSC models.
For simplicity, we have fitted the total emission with the lobe model.
The bottom lines show how the emission of the lobes and the hot spots
would appear if the source were at $z=4$.
}
\label{cyga}
\end{figure}

The radio luminosity of the hot spots of Cygnus A 
is a factor 5--10 smaller than the radio luminosity of its lobes
(Wilson, Young \& Shopbell 2000; see Fig. \ref{cyga}). 
The detection of the hot--spots in X--ray by the Chandra satellite 
allowed Wilson, Young \& Shopbell (2000) to estimate the magnetic field
of the hot spot, $B\sim 1.5\times 10^{-4}$ G, a value close to equipartition
with the emitting electrons.
The corresponding magnetic energy density is $\sim$2,000 times larger than the CMB
energy density at the redshift $z=0.056$ of Cygnus A.
The magnetic field in equipartition with the CMB energy density, $B_{\rm CMB}$ is
\begin{eqnarray} 
B_{\rm CMB}\, &=& \,3.24\times10^{-6}\, \Gamma \,(1+z)^2 \nonumber \\
\, &=& \,8.1\times 10^{-5}\, \Gamma \, \left( { 1+z\over 5} \right)^2 {\rm G}
\label{bcmb}
\end{eqnarray}
where the $\Gamma$ factor accounts for the possibility to have a relativistically moving
source, that would see the CMB energy density amplified by a factor $\Gamma^2$.
If the magnetic field of the hot spots remains $B\sim 10^{-4}$ G at any redshift,
the corresponding magnetic energy density is larger than the radiation energy
density of the CMB up to $z\sim 4.5$.
Consequently, below this redshift, the effects of the CMB on the radio emission
of the hot spots are weak.

\begin{figure}
\vskip -0.6 cm
\hskip -0.2 cm
\psfig{file=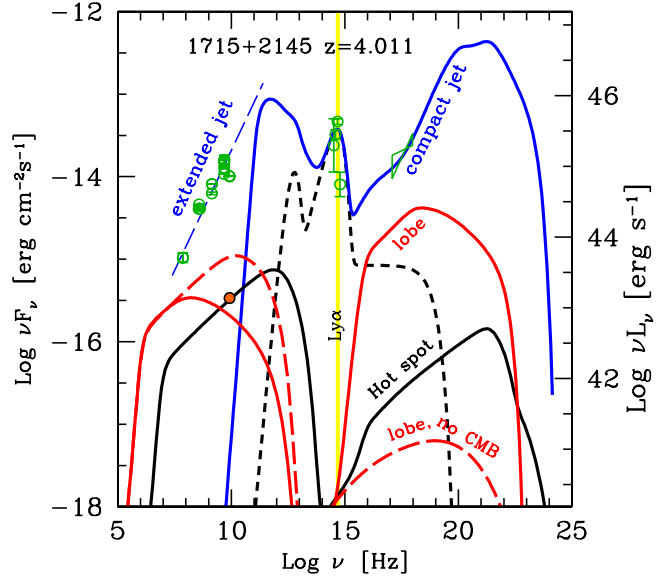,width=9cm,height=9.cm }
\vskip -0.7 cm
\caption{
SED of 1715+2145, together with the models for the different
components of the emission:
(i) the thermal component, namely the accretion disk, the torus and the X--ray corona
(short dashed black line);
(ii) the emission from the compact and relativistically beamed jet (the blue solid line
is the sum of the jet and the thermal components),
accounting for most of the emission, but not for the radio flux, that must be produced
by more extended regions (but still relativistic) of the jet, forming the flat
radio spectrum;
(iii) the emission from the hot spots (solid black line) assumed to have a radius of
$R_{\rm spot}$ = 1.5 kpc, as measured by VLA (Gobeille et al. 2014);
(iv) the lobe emission (solid red line), assuming that the power injected throughout the lobe in the form 
of relativistic electrons is  $P_{\rm e, lobe}=0.1 P_{\rm jet}$. 
Magnetic and electron lobe energies are in equipartition.
For illustration, the  dashed red line shows the lobe emission 
for the same $P_{\rm e, lobe}$ and magnetic field, but now neglecting the effects of CMB photons. 
The long--dashed blue line is not a fitting model, but simply a line to guide the eye.
See Table \ref{para4c} for the adopted parameters.
Data points are from the ASDC archive, the bow---tie in X--rays is from
Wu et al. (2013).
The orange circle corresponds to 4 mJy at 8.4 GHz, and is the flux observed by the two
hot spots.
The vertical yellow line flags the position of the Ly$\alpha$ line.
Above this line the flux is severely absorbed by the intervening matter,
and we do not consider the corresponding data in the fit.
(see for illustration the case of 1026+2542 in Fig. \ref{sed}).
}
\label{1715}
\end{figure}

To model the emission of both the hot spot and the lobe we have followed 
the prescriptions of Ghisellini et al. (2014), that will be summarised in the next
section.
To fit the hot spots of Cygnus A we used
$B=1.1\times 10^{-4}$ G, an injected power of relativistic electrons
$P_{\rm e}=7\times 10^{45}$ erg s$^{-1}$ and a radius of 2 kpc.
The corresponding magnetic energy is $E_{\rm B}=$ erg, a factor 3.8 smaller
than the electron energy $E_{\rm e}=1.6\times 10^{57}$ erg.
For the lobe emission, we have used $B=2\times 10^{-5}$ G, 
same $P_{\rm e}$ and a lobe radius of 58 kpc.
Magnetic and electron energies are the about the same, 
$E_{\rm B}\sim E_{\rm e}= 3\times 10^{59}$ erg.

Fig. \ref{cyga} shows how the SED of the lobe and hot spot emission of Cygnus A
would look like if the source were at $z=4$. 
The shape (and the luminosity) of the hot spot would remain unaltered, since the 
total radiative cooling would be dominated by synchrotron losses even at this redshift.
Its flux would be fainter only because of the larger distance.
On the contrary, the lobe radio spectrum would become much steeper,
and thus fainter than the hot spot emission, due to the enhanced
inverse Compton radiative cooling, now dominating over synchrotron losses.
The enhanced inverse Compton losses produce a correspondingly enhanced
X--ray luminosity.

\vskip 0.2 cm
Among the blazars of our sample, 1715+2145, at $z=4.011$, was
observed by the VLA by Gobeille et al. (2014), who report the detection
of the source at the level of 387 mJy at 8.4 GHz.
Most of this total flux (383 mJy) comes from the unresolved core,
while the remaining 4 mJy are produced by two almost co--aligned sources
of sizes (i.e. diameters) 0.5$\pm$0.1 and 0.4$\pm$0.1 arcsec, corresponding to
3.5$\pm$0.7 and 2.8$\pm$0.7 kpc.
We interpret them as being hot spots, given the similarity of their sizes with the 
size of the hot spots observed in Cygnus A.
For simplicity, we assume that both hot spots have a radius of 1.5 kpc,
and that they contribute 2 mJy each to the observed flux.
The power of each jet is $P_{\rm jet}=1.4\times 10^{47}$ erg s$^{-1}$, and we assumed
that the power in relativistic electrons injected throughout one hot spot
is 10\% of $P_{\rm jet}$. 
To reproduce the observed flux with these assumptions, the magnetic field in the hot
spot must be 190 $\mu$G.
This is not far from equipartition with the emitting electrons, since the corresponding 
magnetic and electron energies are $E_{\rm B}=6\times 10^{56}$ erg
and $E_{\rm e}=1.4\times 10^{57}$ erg.
Fig. \ref{1715} shows the SED of the source together with the assumed models
for the jet, the hot spots and the lobes (for the latter see below).
The hot spot luminosity is a small fraction of the power injected throughout the hot
spot in relativistic electrons:
in 1715+2145 the hot spot emits about $10^{45}$ erg s$^{-1}$, 
to be compared to $P_{\rm e, HS} =1.4\times 10^{46}$ erg s$^{-1}$.
As a consequence, we will assume that the power in relativistic electrons 
injected throughout the hot spot and the lobe
is the same $(P_{\rm e, HS}=P_{\rm e, lobe})$. 
Since the magnetic field of the hot spots is large, their radio emission is unaffected 
by the CMB, and remains the same when switching it off.

\begin{figure*} 
\vskip -0.6 cm
\hskip -1.3 cm
\psfig{file=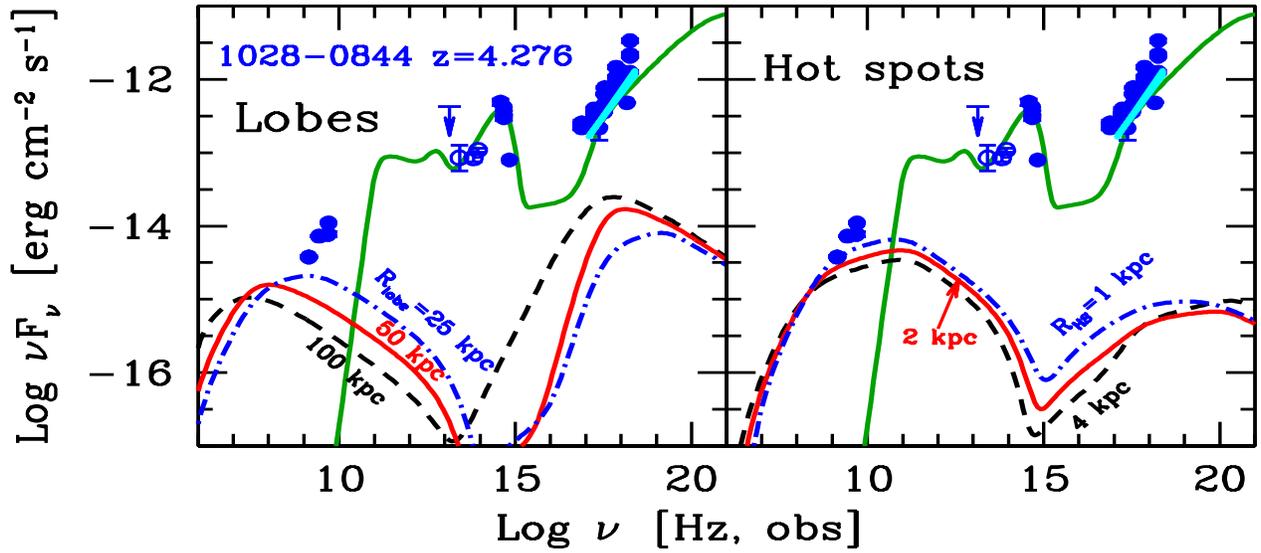,width=19.2cm,height=16cm } 
\vskip -7.3 cm
\caption{
Predicted lobe and hot spot emission from the blazar 1028--0844.
The left panel shows the lobe emission for 
$R_{\rm lobe}=100$ kpc (short dashed black line),
$R_{\rm lobe}=50$ kpc  (solid red), and
$R_{\rm lobe}=25$ kpc (dot--dashed blue).
For all cases we have assumed that the jet deposits in relativistic electrons
the same power, i.e. $0.1 P^{\rm jet}_{\rm e}=2.7\times 10^{46}$ erg s$^{-1}$ for
each lobe and hot spot.
For the right panel the hot spot radius 
$R_{\rm HS}=4$ kpc (dashed black line),
$R_{\rm HS}=2$ kpc  (solid red), and
$R_{\rm HS}=1$ kpc (dot--dashed blue).
The magnetic field is found imposing equipartition with the total electron energy,
both for the lobe and for the hot spot.
The green solid line refers to the jet model (as in Fig. \ref{1028power} and \ref{1028gamma}).
See Table \ref{para4c} for the adopted parameters.
}
\label{1028r}
\end{figure*} 

\subsection{Lobe emission}
\label{lobe}

Theoretical models of jet--powered radio lobes rely on high--resolution 
numerical magneto--hydrodynamic simulations. 
A crucial role in dictating lobe formation and evolution is played by the environment in which 
they develop (see, e.g., Hardcastle \& Krause 2014 for state--of--the--art simulations and references). 
While low--$z$ radio galaxies inhabit well--known astrophysical environments, such as poor galaxy clusters, 
the boundary conditions for the high--$z$ jets we are interested in are poorly known at best.  
As a consequence, we have to rely on numerical work and/or the observed blazar SEDs  themselves to estimate 
some of the fiducial values for modelling the lobe emission.  

Following Ghisellini et al. (2014a), we assume a spherical emitting region of radius $R_{\rm lobe}$, 
which is homogeneously filled with a magnetic field with 
a coherence length--scale $\lambda=10$ kpc (see Carilli \& Taylor 2002; Celotti \& Fabian 2004). 
Relativistic electrons are injected into the lobe with a total
power $P_{\rm e, lobe}$ and a distribution with the same functional form
$Q(\gamma)$ of Eq. \ref{qgamma} (but with parameters different from the jet ones).
We are assuming that the jet power is essentially constant all along its length,
since the radiated power is estimated to be $\sim$10\% of $P_{\rm jet}$
(Nemmen et al. 2012; Ghisellini et al. 2014c).
The steady--state electron energy distribution is found by solving the continuity equation 
at particle crossing time $t_{\rm cross} = (R_{\rm lobe}/c) (1+R_{\rm lobe}/\lambda)$. 

Fig. \ref{1715} shows the SED of 1715+2145 and the lobe model assuming
$P_{\rm lobe}=0.1 P_{\rm jet}=1.4\times 10^{46}$ erg s$^{-1}$, a size $R_{\rm lobe}=50$ kpc,
and equipartition between relativistic electrons and magnetic field, giving $B$=20 $\mu$Gauss for the lobe.
For comparison, we show how the lobe emission would look like by neglecting the interaction of electrons with the CMB. 
The CMB enhances the X--ray emission by $\simeq 2$ orders of magnitude, while dimming 
the radio emission at frequencies $\gsim 100$ MHz.
The lobe emission at lower $\nu$ is instead not CMB--quenched.
This happens because low energy electrons, responsible for the radio flux at $\nu \lsim 100$ MHz,  
do not have enough time to cool, even considering the extra losses due to the interactions with CMB photons.
Their number, obtained from the continuity equation, is not controlled by the cooling rate,
rather it simply equals the injection rate multiplied by the time elapsed from the start of the injection
(see Ghisellini et al. 2014a for full details).

\begin{figure*} 
\vskip -0.7 cm
\hskip -1.3 cm
\psfig{file=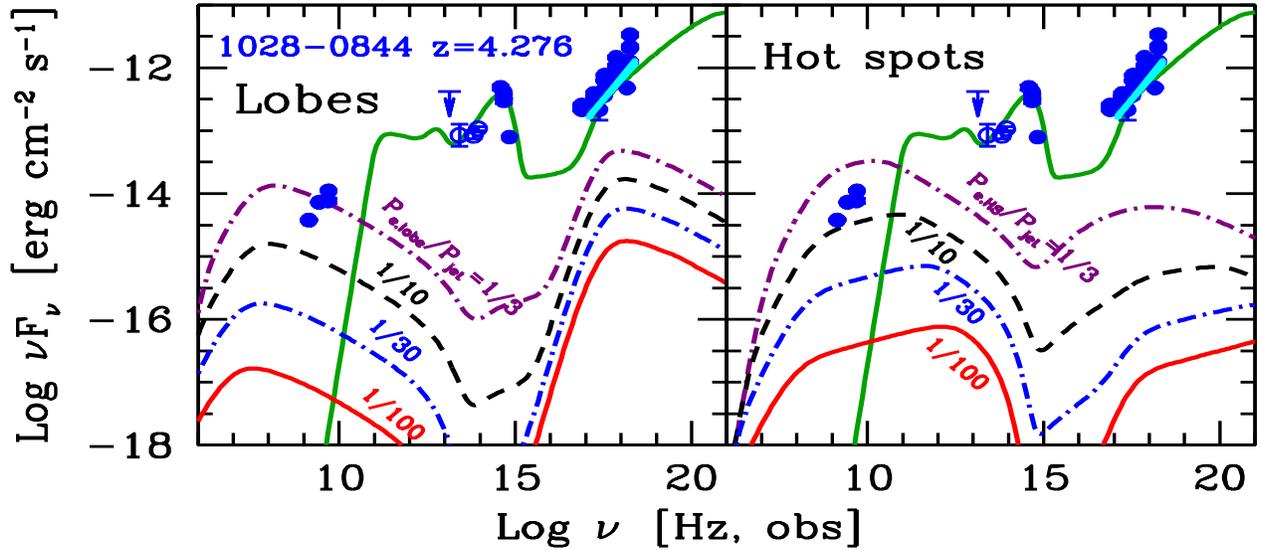,width=19.2cm,height=16cm } 
\vskip -7.3 cm
\caption{
Predicted lobe and hot spot emission from the blazar 1028--0844
as a function of the injected power in relativistic electrons.
For both panels
$P_{\rm e}/P_{\rm jet}=1/3$ (violet dot--dashed line);
1/10 (dashed black line); 1/30 (dot--dashed blue line) and 1/100 (solid red line).
For all cases we have assumed that the lobe radius is 50 kpc and the hot spot radius is 2 kpc.
The magnetic field is found imposing equipartition with the total electron energy.
A larger $P_{\rm e}$ implies a larger equipartition magnetic field and this in
turn implies that the synchrotron luminosity increases more than the inverse Compton one.
See Table \ref{para4c} for the adopted parameters.
}
\label{1028power}
\end{figure*} 

\begin{figure*} 
\vskip -0.7 cm
\psfig{file=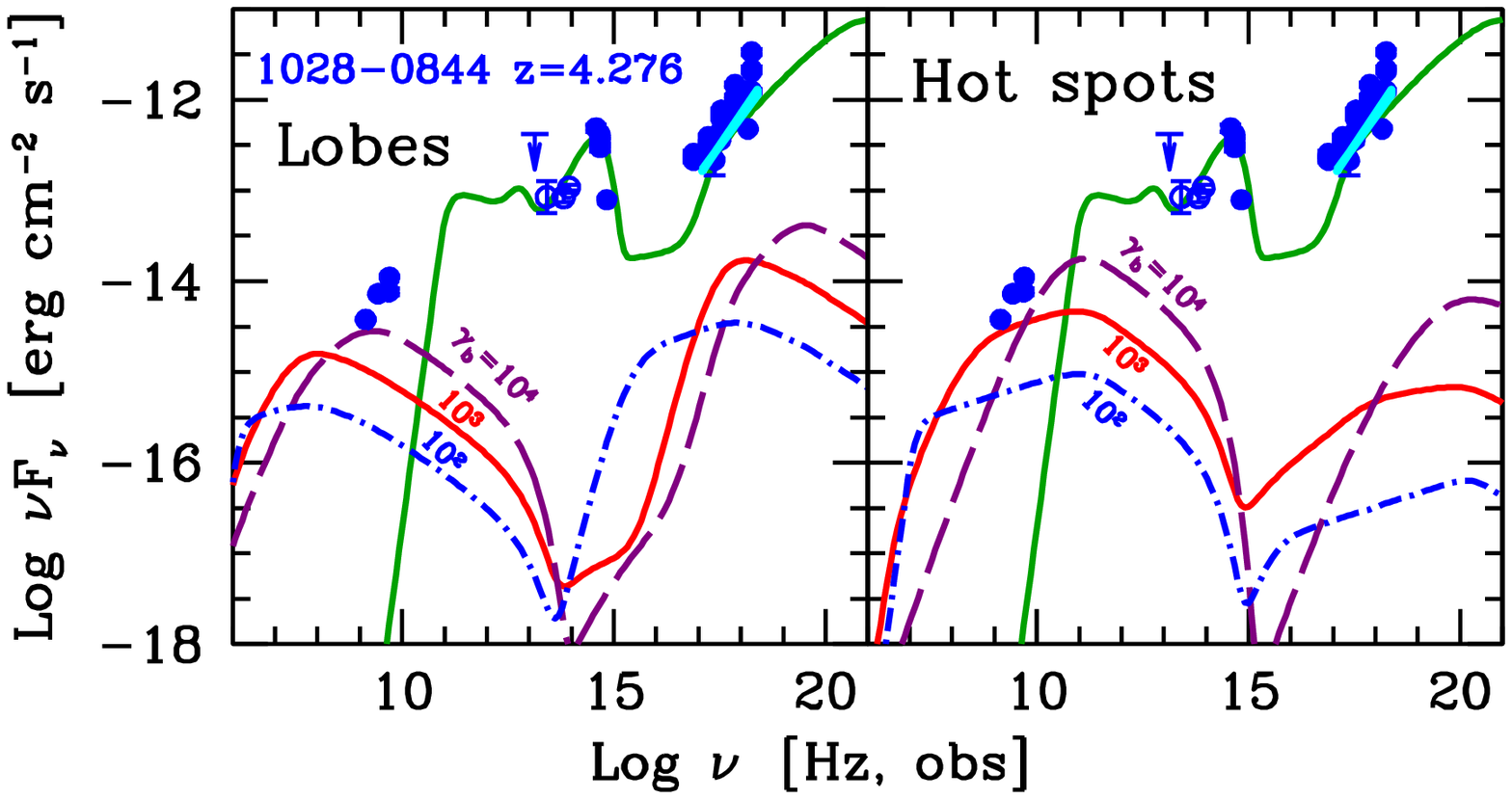,width=19.2cm,height=16cm } 
\vskip -7.6 cm
\hskip -1.3 cm
\caption{
Predicted lobe and hot spot emission from the blazar 1028--0844
as a function of the break energy of the injected electrons $\gamma_{\rm b}$
(see Eq. 2).
For both panels:
$\gamma_{\rm b}=10^{4}$ (long dashed violet line);
$\gamma_{\rm b}=3\times 10^{3}$ (solid red line);
$\gamma_{\rm b}=3\times 10^{2}$ (dot--dashed blue line).
For all cases we have assumed that the lobe radius is 50 kpc and the hot spot radius is 2 kpc.
The magnetic field is found imposing equipartition with the total electron energy.
See Table \ref{para4c} for the adopted parameters. 
}
\label{1028gamma}
\end{figure*} 

\subsubsection{Constraints on the physical parameters of the hot spot and lobe}

In this part of the section we discuss the constraints we can put on the 
physical parameters of the hot spot and lobe, and how the predicted spectrum depends upon the assumed values. 
To this aim we consider one particular blazar in our sample, namely 1028--0844. 
In all tested models the magnetic field is estimated by requiring  equipartition 
with the electron total energy. 
Equipartition is actually supported by observations of classical double FR II radio--galaxies in 
the radio and X--ray bands, and appears as an educated guess 
(Wilson et al. 2000; Croston et al. 2004, 2005; Belsole et al. 2004). 

\vskip 0.2 cm
\noindent
{\it Size ---}
The radius of the hot spots of the nearby radio--galaxy Cygnus A and of the distant blazars
1715+2145 is close to 2 kpc for both sources.
In Fig. \ref{1028r} (right panel) we show the emission assuming $R_{\rm HS}=$1, 2 and 4 kpc,
always requiring equipartition between the magnetic and electron energy densities.
One can see that the hot spot emission is not sensitive to the changes in size.
The assumption $P_{\rm e, HS}=0.1 P_{\rm jet}$, for this particular source, leads to an 
overestimation of the observed radio spectrum, once the hot spot emission is summed to the
flat radio emission of outer parts of the jet.
The synchrotron luminosity dominates over the inverse Compton one, and their sum is a small fraction
of $P_{\rm e, HS}$.

At low redshift, radio lobes form in the overdense environment appropriate for a galaxy cluster. 
However, at $z\gsim 4$ clusters have to form yet, and we can argue that lobes of radio galaxies 
develop within a medium at (or close to) the mean cosmic density. 
It turns out that at $z\simeq 4$ the mean cosmic density is comparable to the typical density of the 
intra--cluster medium $\simeq 1$ Mpc away from the center of a typical formed cluster at $z\lsim1$. 
Because of this coincidence, we then expect that the lobe sizes of high--redshift blazars  are not 
very different from those of very powerful radio sources residing in low--$z$ virialized clusters
(although some additional effect may cause the apparent lack of hot spots and lobes at distances
larger than 100 kpc for FR II radio--galaxies at $z\gsim$1 found by van Velzen, Falcke \& K\"ording 2015).

The left panel of Fig. \ref{1028r} shows the predicted lobe emission assuming different values of $R_{\rm lobe}$. 
Different values of $R_{\rm lobe}$ lead to different values of the equipartition magnetic field 
(see the values in Tab. \ref{para4c}) and hence of the self--absorption synchrotron frequency.
Therefore, although the power injected in the lobes is fixed, the synchrotron luminosity
gets smaller for larger $R_{\rm lobe}$ as this corresponds to a smaller magnetic field. 
The X--ray luminosity is instead nearly independent upon $R_{\rm lobe}$, because 
in the fast cooling regime (i.e., when most electrons with $\gamma>\gamma_{\rm b}$ cool radiatively) 
the emitted power balances the (assumed constant) injected power. 
Fig. \ref{1028r} indicates that, under our assumptions, $R_{\rm lobe}$ has to be $\gsim $25 kpc to 
keep model predictions consistent with the observed SED.

\vskip 0.2 cm
\noindent
{\it Injected power in relativistic electrons ---}
We assume that the power injected in relativistic electrons throughout the hot spot an the lobe
lobe is a fraction of the total jet power $P_{\rm jet}$.
The value of this fraction is basically unknown, 
but we can set $P_{\rm e, HS}=P_{\rm e, lobe}$ since radiative losses are not important in the hot spot.

If magnetic field, proton energy and electron energy are in equipartition 
and in addition we account for the $pV$ work of the lobe, then $\sim$10\% appears a fair guess.
On the other hand, recent numerical work by Hardcastle \& Krause (2014; see their Fig. 5), 
suggests a value close to 1\% (characteristic of an environment where hot protons dominate the
global energetics). 

The right panel of Fig. \ref{1028power} shows how the emission from the hot spots changes
by  changing $P_{\rm e HS}$.
For $P_{\rm e, HS}=10^{-2} P_{\rm jet}$ the corresponding magnetic field is relatively small,
and most electrons do not cool while crossing the hot spot.
As a consequence the particle distribution does not steepen because of cooling, and the spectrum is hard. 
As $P_{\rm e, HS}$ increases, also the equipartition magnetic field increases (see Table \ref{para4c}), 
together with the cooling rate, inducing a steepening in the particle distribution and 
in the produced  spectrum.

The left panel of Fig. \ref{1028power} illustrates the effect of changing $P_{\rm e, lobe}$ while 
keeping $R_{\rm lobe}=50$ kpc and the magnetic field in equipartition with the total electron energy. 
This forces the synchrotron luminosity to increase as approximatively $\propto P_{\rm e, lobe}^2$, since 
a larger $P_{\rm e, lobe}$ implies a higher electron density {\it and} a higher value of the equipartition 
magnetic field.
By comparing the different models in Fig. \ref{1028power}, we conclude that, 
unless $R_{\rm lobe}\gg 50$ kpc, assuming  
$P_{\rm e, lobe}$ in excess of 10\% $P_{\rm jet}$ leads to an overestimate of the radio emission.

\vskip 0.2 cm
\noindent
{\it Break energy of the injected electrons ---}
Fig. \ref{1028gamma} illustrates the effects of varying the electron energy at the break of the 
energy distribution (Eq. \ref{qgamma}).
An increase of $\gamma_b$ corresponds to an increase in the power injected
in high energy electrons (which cool more efficiently). 
This translates into a higher bolometric luminosity and a reduced total energy.
This is true both for the hot spot and the lobe.
Higher values of $\gamma_{\rm b}$ also correspond to a larger Compton dominance, with
the inverse Compton spectrum peaking at larger frequencies.

\vskip 0.2 cm
\noindent
We can now summarise our fiducial values for 
the hot spot and the lobe parameters. 
We will assume all sources have hot spots of size $R_{\rm HS}=$2 kpc, and that
$P_{\rm e, HS} = P_{\rm e, lobe}=0.1P_{\rm jet}$ and  $P_{\rm e, HS} = P_{\rm e, lobe}=0.01P_{\rm jet}$.
We will assume equipartition between the magnetic field 
and the total electron energy, while for the injected particle distribution (i.e. Eq. \ref{qgamma}) we will take  
power--law indices $s_1=-1$ and $s_2=2.7$, and break and maximum Lorentz factor 
$\gamma_{\rm b}=10^3$ and $\gamma_{\rm max}=10^6$, respectively. 
As mentioned, the steep value of $s_2$ makes the exact value of 
$\gamma_{\rm max}$ not critical.

\begin{figure}
\vskip -0.7 cm
\hskip -0.3 cm
\psfig{file=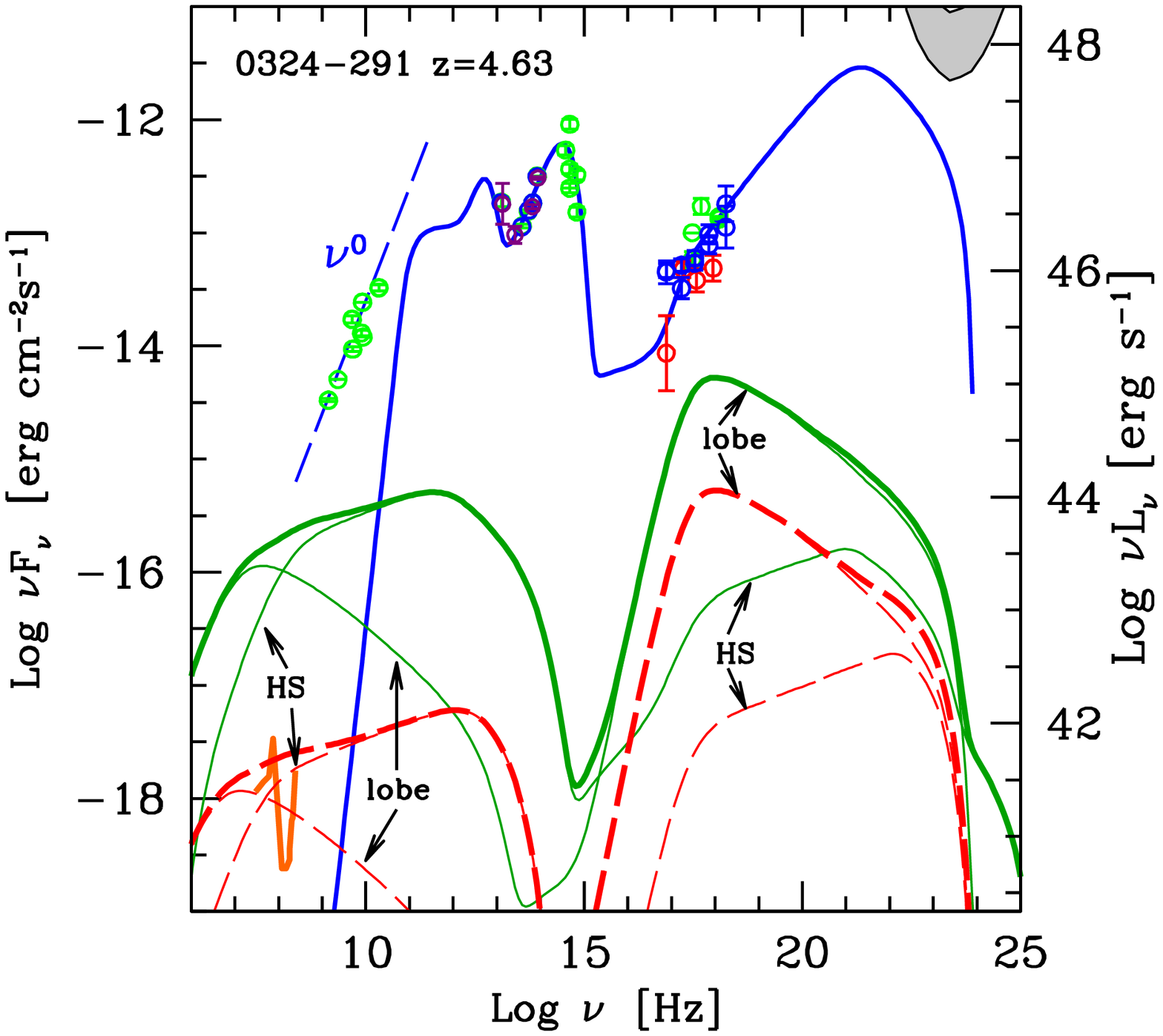,width=9cm,height=9.cm }
\vskip -0.7 cm
\caption{
SED of the blazar 0324--291. 
The blue solid line is the model for the sum of jet emission and thermal components 
(see Fig. \ref{1224}). 
For the hot spots and the lobes we show two possible models:
one with 0.1$P_{\rm jet}$ (thin green lines)
and the other with 0.01$P_{\rm jet}$ (thin red lines).
We show separately the contribution of the hot spots and the lobes, as labelled,
while the thick green and red lines are the corresponding sums.
The radius of the hot spots and the lobes are assumed to be $R_{\rm HS}=$2 kpc 
and $R_{\rm lobe}=$50 kpc.
The straight dashed blue line interpolating the radio spectrum is not a fit,
but only a guide to the eye.
The hatched grey area in the upper right corner shows the sensitivity of 
{\it Fermi}/LAT (5$\sigma$) after 5 years of observing time.
We also report (orange line) the sensitivity of LOFAR.
}
\label{0324}
\end{figure}

\begin{figure}{7}
\vskip -0.6 cm
\psfig{file=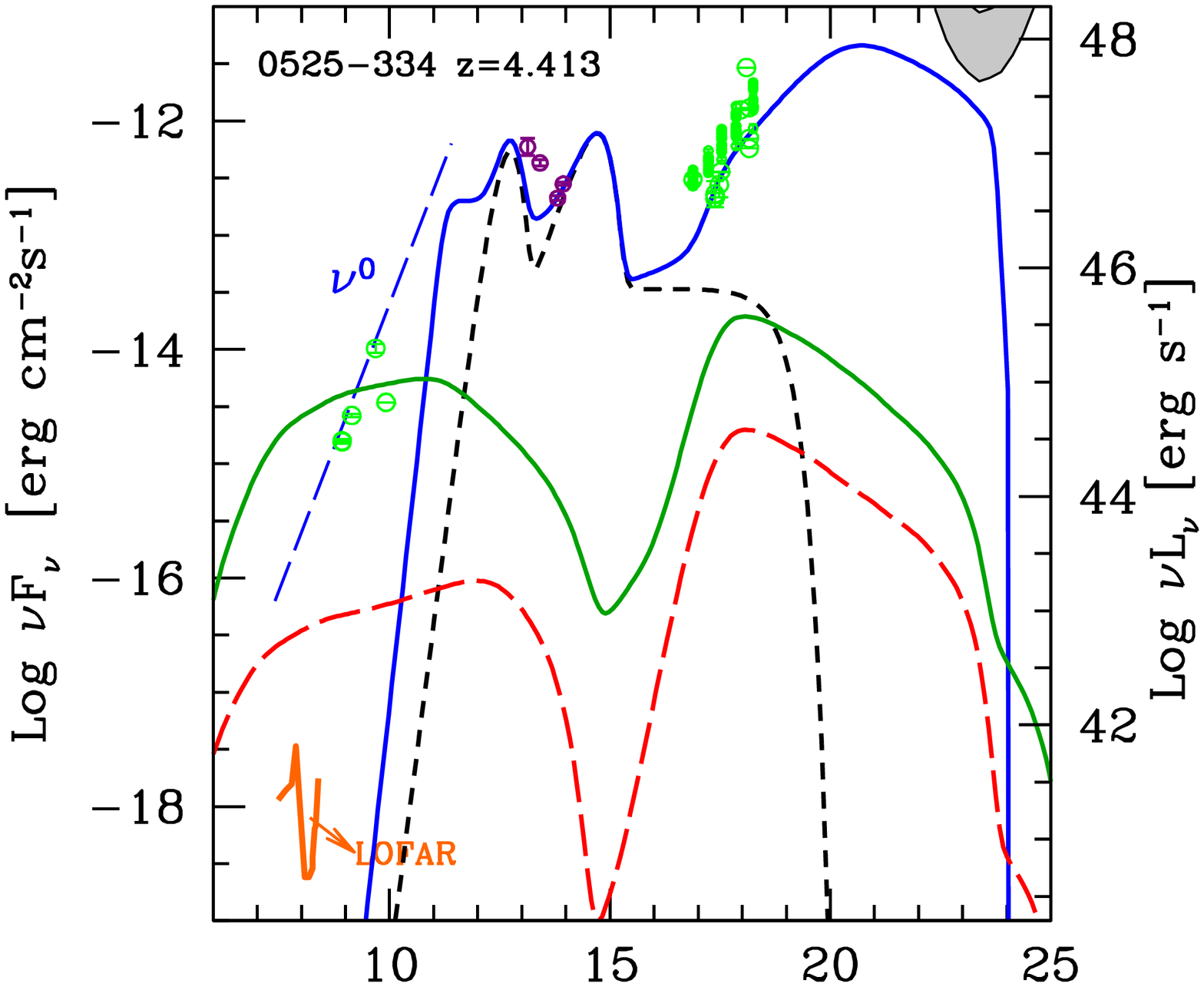,width=9cm,height=7.cm }
\vskip -1.3 cm
\psfig{file=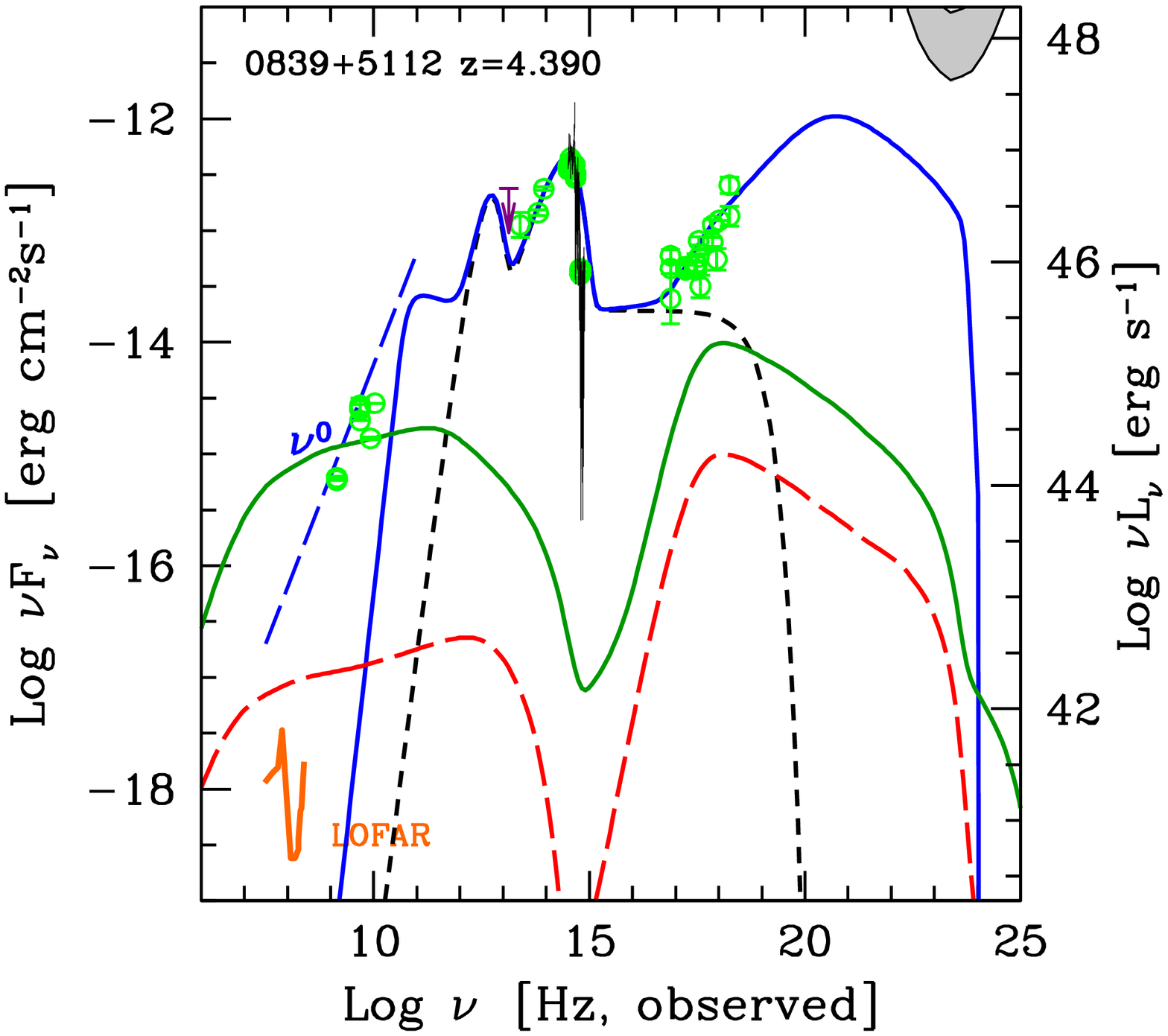,width=9cm,height=7.cm }
\vskip -0.7 cm
\caption{
SED of the $z>4$ blazars.
As in Fig. \ref{0324}, but showing only the sum of the hot spot and lobe components.
For 0839+5112 the black solid line shows the SDSS spectrum dereddened for Galactic absorption.
}
\label{sed}
\end{figure}
\setcounter{figure}{7}
\begin{figure}
\vskip -0.6 cm
\psfig{file=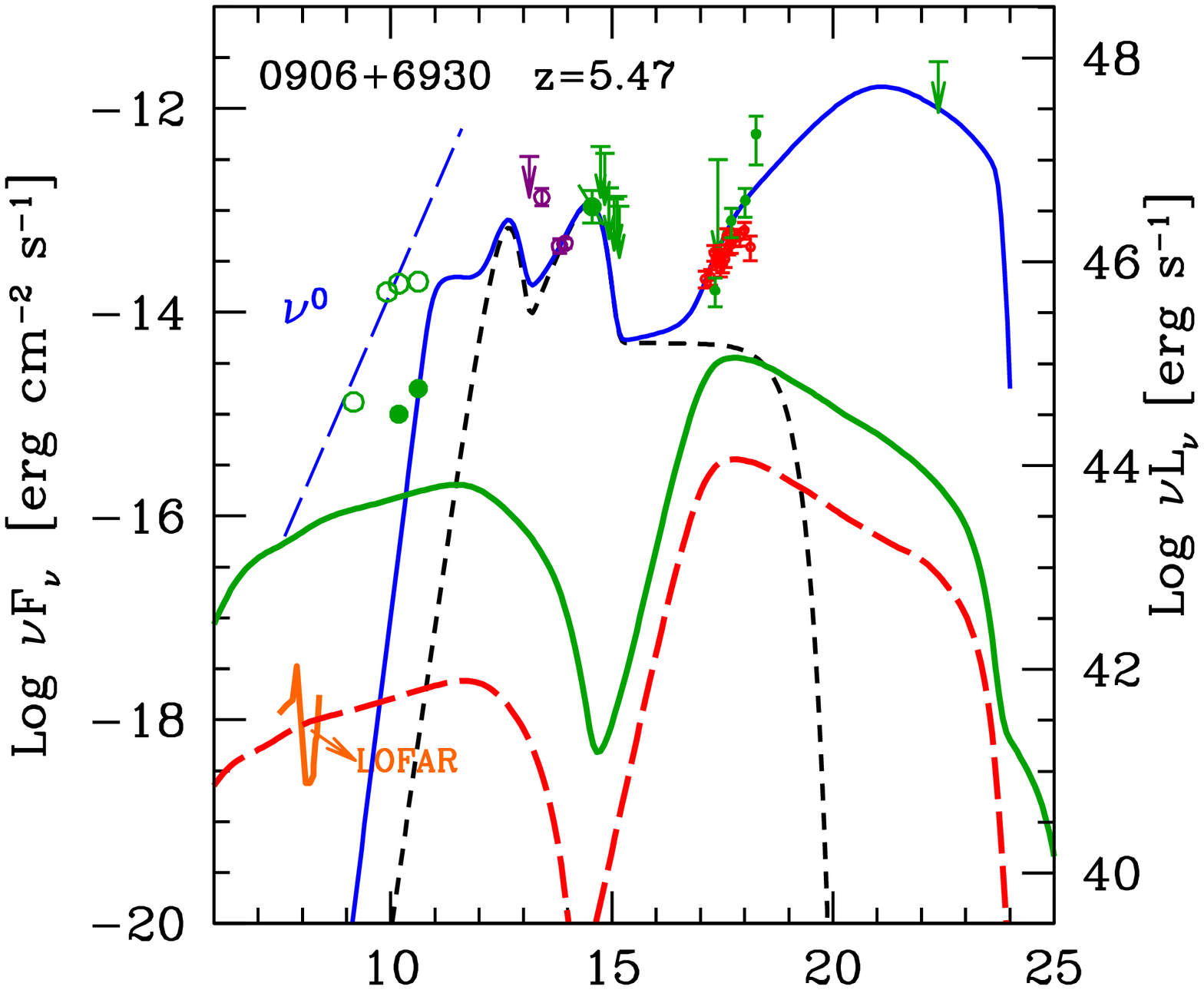,width=9cm,height=7.cm }
\vskip -1.3 cm
\psfig{file=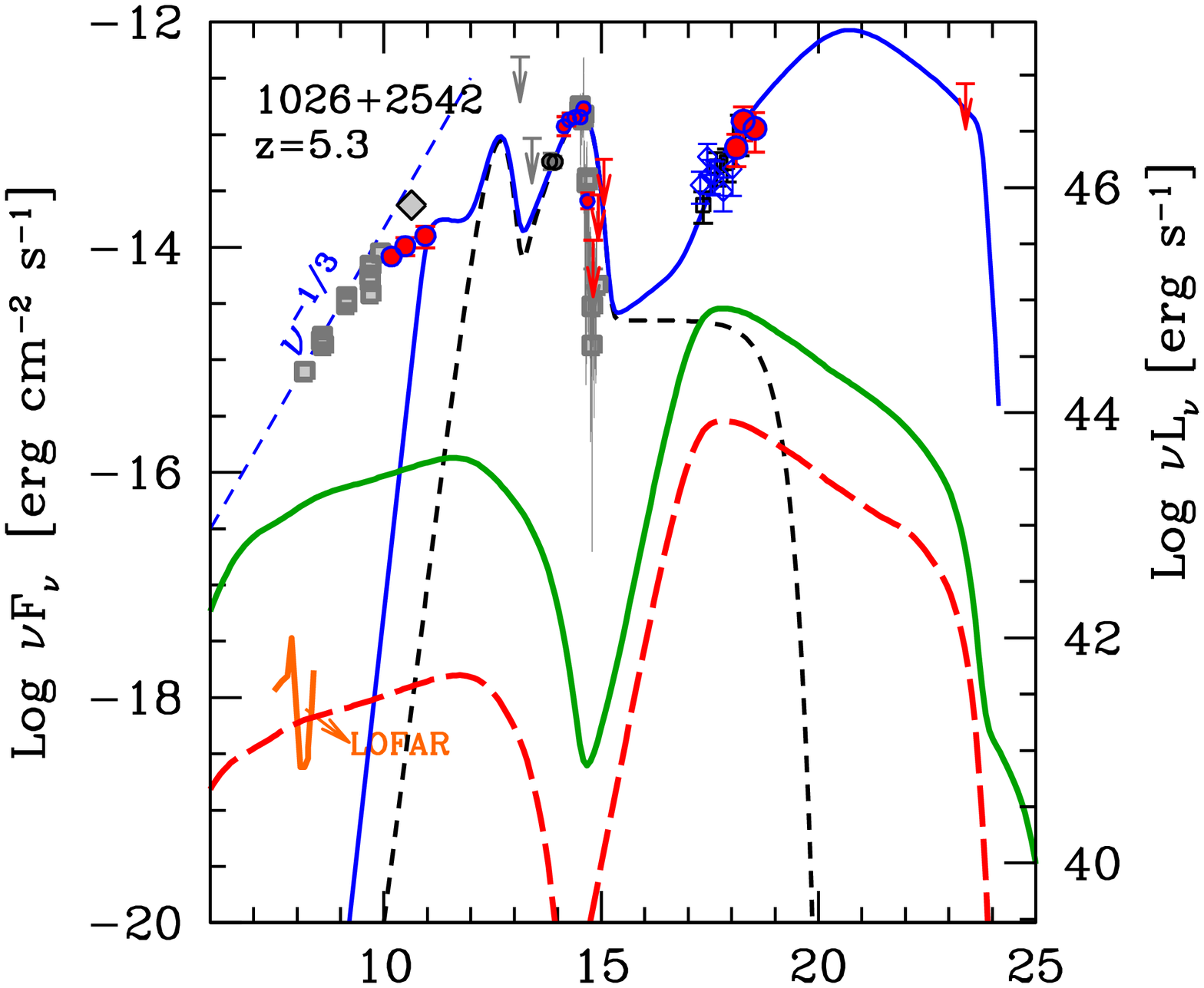,width=9cm,height=7.cm }
\vskip -1.3 cm
\psfig{file=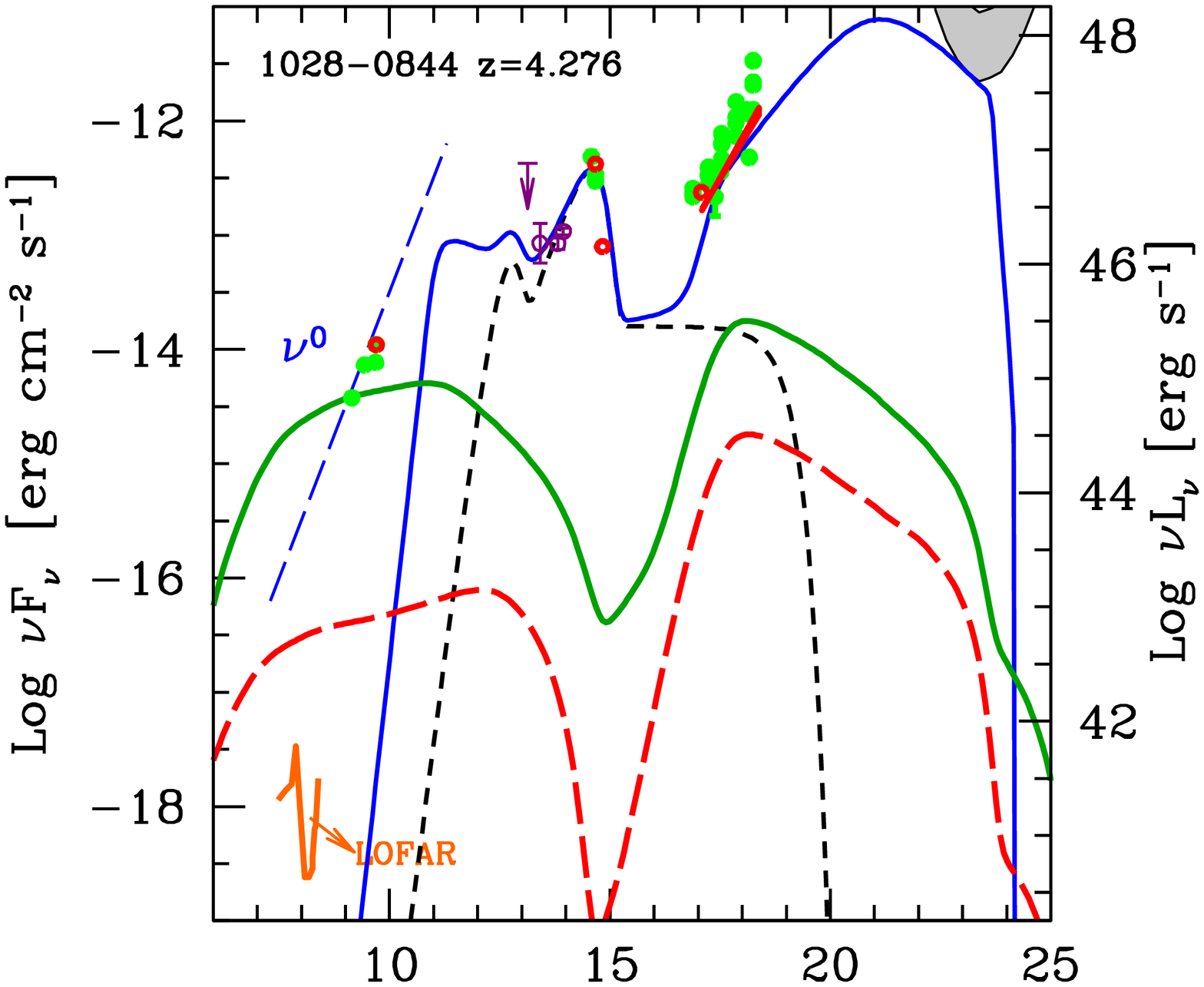,width=9cm,height=7.cm }
\vskip -1.3 cm
\psfig{file=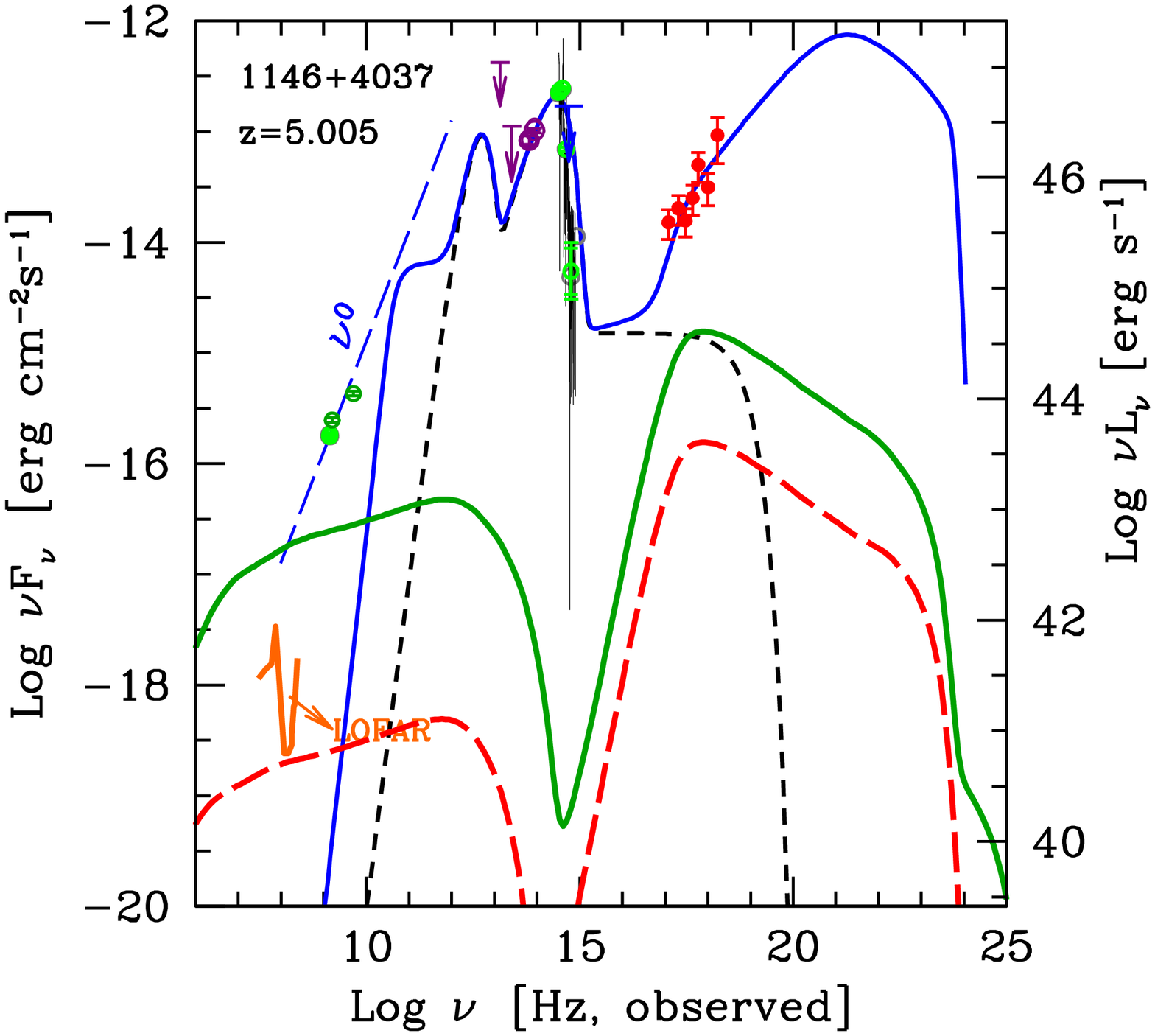,width=9cm,height=7cm }
\vskip -0.7 cm
\caption{{\it continue.}
SED of the $z>4$ blazars.
As in Fig. \ref{0324}, but showing only the sum of the hot spot and lobe components.
}
\label{sed}
\end{figure}

\begin{table*} 
\centering
\begin{tabular}{l l l l l l l l l l l l l l l l l}
\hline
\hline
Name &$z$ &$R_{\rm diss}$ &$R_{\rm BLR}$ &$P^\prime_{\rm e, jet, 45}$  &$B$ &$\Gamma$ &$\theta_{\rm V}$  
  &$\gamma_{\rm b}$ &$\gamma_{\rm max}$ &$s_1$ &$s_2$  &$\log P_{\rm r}$  &$\log P_{\rm jet}$ \\
~[1] &[2] &[3] &[4] &[5] &[6] &[7] &[8] &[9] &[10] &[11] &[12] &[13] &[14]    \\
\hline   
0324 --2918  &4.630  &1200 &1140 &0.02  &2.4 &13   &3 &150 &2e3 &--1 &2.8 &46.0 &46.9  \\ 
0525 --3343  &4.413  &630  &1643 &0.06  &2.9 &15   &3 &50  &3e3 &0   &2.4 &46.1 &47.5  \\ 
0839 +5112   &4.390  &1470 &1336 &0.04  &1.2 &13   &4 &90  &4e3 &0   &2.6 &45.8 &47.2  \\ 
0906 +6930   &5.47   &630  &822  &0.02  &1.8 &13   &3 &100 &3e3 &0   &2.5 &46.0 &46.8  \\ 
1026 +2542   &5.304  &504  &916  &0.01  &2.3 &13   &3 &70  &4e3 &0   &2.6 &45.7 &46.7  \\ 
1028 --0844  &4.276  &840  &1095 &0.08  &1.4 &15   &3 &100 &4e3 &0   &2.6 &46.2 &47.4  \\ 
1146 +4037   &5.005  &900  &1006 &7e--3 &1.4 &13   &3 &130 &3e3 &0   &2.6 &45.6 &46.4  \\ 
1253 --4059  &4.460  &480  &671  &0.02  &1.8 &13   &3 &100 &3e3 &1   &2.5 &45.5 &47.2  \\ 
1309 +5733   &4.268  &1200 &756  &3e--3 &0.5 &12   &3 &20  &3e3 &0   &2.7 &44.6 &46.2     \\             
1325 +1123   &4.412  &900  &739  &9e--3 &1.9 &12   &3 &250 &3e3 &0   &2.6 &45.5 &46.3   \\
1420 +1205   &4.034  &360  &725  &6e--3 &2.6 &13   &3 &120 &3e3 &1   &2.5 &45.5 &46.7 \\ 
1430 +4204   &4.715  &540  &1112 &0.1   &1.7 &14   &3 &40  &3e3 &0   &2.6 &46.8 &47.9 \\ 
1510 +5702   &4.309  &293  &636  &0.04  &2.6 &15   &3 &60  &4e3 &0.5 &2.6 &46.4 &47.5 \\ 
1715 +2145   &4.011  &396  &233 &0.013 &0.65 &15   &3 &220 &3e3 &1.5 &3.2 &45.3 &47.2  \\
2134 --0419  &4.346  &432  &972  &7e--3 &2.9 &13   &3 &70  &4e3 &0   &2.6 &45.5 &46.6  \\ 
2220 +0025   &4.205  &360  &671  &3e--3 &2.4 &13   &3 &120 &3e3 &1   &2.2 &45.2 &46.4  \\ 
\hline
\hline 
\end{tabular}
\vskip 0.4 true cm
\caption{
Adopted parameters for the jet models shown in Fig. \ref{sed}.
Col. [1]: name; 
Col. [2]: redshift;
Col. [3]: distance of the dissipation region from the black hole, in units of $10^{15}$ cm;
Col. [4]: size of the BLR, in units of $10^{15}$ cm;
Col. [5]: power injected in the jet in relativistic electrons, calculated in the comoving 
frame, in units of $10^{45}$ erg s$^{-1}$;
Col. [6]: magnetic field in G;  
Col. [7]: bulk Lorentz factor;
Col. [8]: viewing angle in degrees;
Col. [9] and Col. [10]: break amd maximum Lorenz factor of the injected electron distribution;
Col. [11] and Col. [12]: slopes of the injected electron distribution; 
Col. [13]: logarithm of jet power in the form of radiation, in erg s$^{-1}$;
Col. [14]: logarithm of the total kinetic plus magnetic jet power, in erg s$^{-1}$.
The values of the powers and the energetics refer to {\it one} jet.
}
\label{para}
\end{table*}
\begin{table} 
\centering
\begin{tabular}{l l l l l l l l l l l l l l l l l}
\hline
\hline
Name    &$\log P_{\rm e}$ &$B_{\rm HS}$  &$\log E_{\rm B, HS}$ &$B_{\rm lobe}$  &$\log E_{\rm B, lobe}$   \\ 
~[1] &[2] &[3] &[4] &[5] &[6] \\
\hline   
0324 --2918    &45.9 &172  &57.1 &13.5 &59.1    \\  
               &44.9 &52.2 &56.0 &13.5 &58.1   \\
0525 --3343    &46.5 &319  &57.6 &26   &59.6   \\  
               &45.5 &105  &56.7 &8.2  &58.6   \\ 
0839 +5112     &46.2 &233  &57.3 &18.2 &59.3   \\   
               &45.2 &71   &56.3 &5.8  &58.3   \\
0906 +6930     &45.8 &151  &56.9 &11.9 &59.9   \\   
               &44.8 &47   &55.9 &3.77 &57.9   \\                
1026 +2542     &45.7 &130  &56.8 &10.5 &58.8   \\    
               &44.7 &41.2 &55.8 &3.32 &57.8   \\                           
1028 --0844    &46.4 &310  &57.6 &24.5 &59.6   \\    
               &45.4 &97   &56.6 &7.8  &58.6   \\  
1146 +4037     &45.4 &95   &56.6 &7.53 &58.5   \\   
               &44.4 &29.3 &55.5 &2.4  &57.5   \\  
1253 --4059    &46.2 &233  &57.3 &18.1 &59.3   \\   
               &45.2 &71.4 &56.3 &5.8  &58.3   \\
1309 +5733     &45.0 &56.8 &56.1 &4.8  &58.2 \\
               &44.0 &16.5 &55.0 &1.51 &57.1 \\
1325 +1123     &45.3 &82   &56.5 &6.7  &58.4 \\
               &44.3 &24.2 &55.4 &2.12 &57.4 \\
1420 +1205     &45.7 &138  &56.9 &11.1 &58.9   \\    
               &44.7 &39   &55.8 &3.3  &57.8   \\               
1430 +4204     &46.9 &510  &58.0 &41   &60.0   \\   
               &45.9 &169  &57.0 &13.2 &59.0   \\   
1510 +5702     &46.5 &335  &57.6 &26.8 &59.6  \\  
               &45.5 &106  &56.6 &8.5  &58.6  \\ 
1715 +2145$^*$ &46.1 &190  &56.8 &20.3 &59.4 \\
2134 --0419    &45.6 &115  &56.7 &9.35 &58.7  \\   
               &44.5 &34.3 &55.7 &2.95 &57.7  \\              
2220 +0025     &45.4 &89   &56.5 &7.23 &58.5  \\   
               &44.4 &25.5 &55.4 &2.3  &57.5  \\  
\hline
\hline 
\end{tabular}
\vskip 0.4 true cm
\caption{
Adopted parameters for the hot spot and lobe models shown in Fig. \ref{0324} and Fig. \ref{sed}.
For each source, the first raw corresponds to $P_{\rm e}/P_{\rm jet}=0.1$,
while the second raw corresponds to $P_{\rm e}/P_{\rm jet}=10^{-2}$.
Col. [1]: name; 
Col. [2]: logarithm of the power injected throughout the hot spots and lobes in relativistic electrons in erg s$^{-1}$;
Col. [3]: magnetic field of the hot spot in $\mu$G;
Col. [4]: logarithm of the energy in magnetic field contained in the hot spot, in erg;
Col. [5]: magnetic field of the lobe in $\mu$G;
Col. [6]: logarithm of the energy in magnetic field contained in the lobe, in erg;
All sources have $R_{\rm HS}= 2$ kpc, $R_{\rm lobe}= 50$ kpc,
$\gamma_{\rm b}=10^3$, $\gamma_{\rm max}=10^6$, and
slopes of the injected electron distribution $s_1=-1$ and $s_2=2.7$.
The values of the powers and the energetics refer to {\it one} jet and {\it one} hot spot and lobe,
while the lobe flux shown in the figures corresponds to {\it two} hot spots and lobes.
$^*$: for this source the hot spot is detected, so we have used the observed $R_{\rm HS}=$1.5 kpc.
The magnetic field is found without invoking equipartition, and  the electron energy of the hot spot
is $\log(E_{\rm e, HS}/\rm erg)=57.1$.
}
\label{paralobe}
\end{table}

\section{Results}

The jet, hot spot and lobe modelling described in the previous 
sections are applied to the broad band SEDs of our sample of high--$z$ blazars. 
In Fig. \ref{0324} and Fig. \ref{sed} we show the observed SEDs,   
along with the corresponding  jet and hot spot+lobe emission models. 
Models of hot spot
and lobe emission take into account the CMB when solving for the electron radiative
cooling, and for the resulting energy distribution and emitted spectrum.

The parameters required to describe the SED from mm to hard X--ray bands are listed in 
Table \ref{para} and Table \ref{paralobe}, 
while the derived black hole masses and 
accretion disc luminosities are reported in Table \ref{sample}.  
Quite remarkably, the resulting best--fit parameters 
of the jet are very similar to what found for the much larger sample of 
{\it Fermi}/LAT blazars analysed by Ghisellini \& Tavecchio (2015) and Ghisellini et al. (2014c), though 
none of our high--$z$ blazars has been actually detected by {\it Fermi}/LAT at $\gamma$--ray energies.

Typical sizes for the dissipation region are in the range $(3-15)\times 10^{17}$ cm, which lies  
within the BLR (except in the case of 0839+5112, for which $R_{\rm diss}$ is slightly larger).
The jet magnetic field is found to be $\sim 1-2$ G, the bulk jet Lorentz factor $\Gamma\sim 13-15$, 
and the jet power $46.4< \log (P_{\rm jet}/{\rm erg \, s^{-1}})< 47.9$.

Once that data are fit to the composite jet/disk/torus model, we can estimate the physical 
parameters of the hot spot and lobe. 
Assuming  $P_{\rm e}/P_{\rm jet}=0.1$, the equipartition magnetic field is estimated to be
between 82 and and 510 $\mu$G for the hot spots and between 7 and 41 $\mu$G for the lobes,
while the total energies range from 
$\log(E_{\rm B}/{\rm erg})=\log (E_{\rm e}/{\rm erg})= 56.1$ to 58 for the hot spots and
$\log(E_{\rm B}/{\rm erg})=\log (E_{\rm e}/{\rm erg})= 58.5$ to 60 for the lobes.
For the  $P_{\rm e, lobe}/P_{\rm jet}=10^{-2}$ case we find
$B_{\rm HS}$ between 16.5 and 169 $\mu{\rm G}$, 
$B_{\rm lobe}$ between 2.3 and 13 $\mu{\rm G}$, 
and $\log(E_{\rm B}/{\rm erg})=\log (E_{\rm e}/{\rm erg})$ between 55 and 57 for the hot spots
and between 57.5 and 59 for the lobes.

The calculated hot spot+lobe emission, for $P_{\rm e, lobe}/P_{\rm jet}=0.1$, sometimes
exceeds the observed data, requiring that less power is
deposited into the hot spots and the lobes.
For $P_{\rm e, lobe}/P_{\rm jet}=10^{-2}$ we have consistency
with the existing radio data for all blazars.

\setcounter{figure}{7}
\begin{figure}
\vskip -0.6 cm
\psfig{file=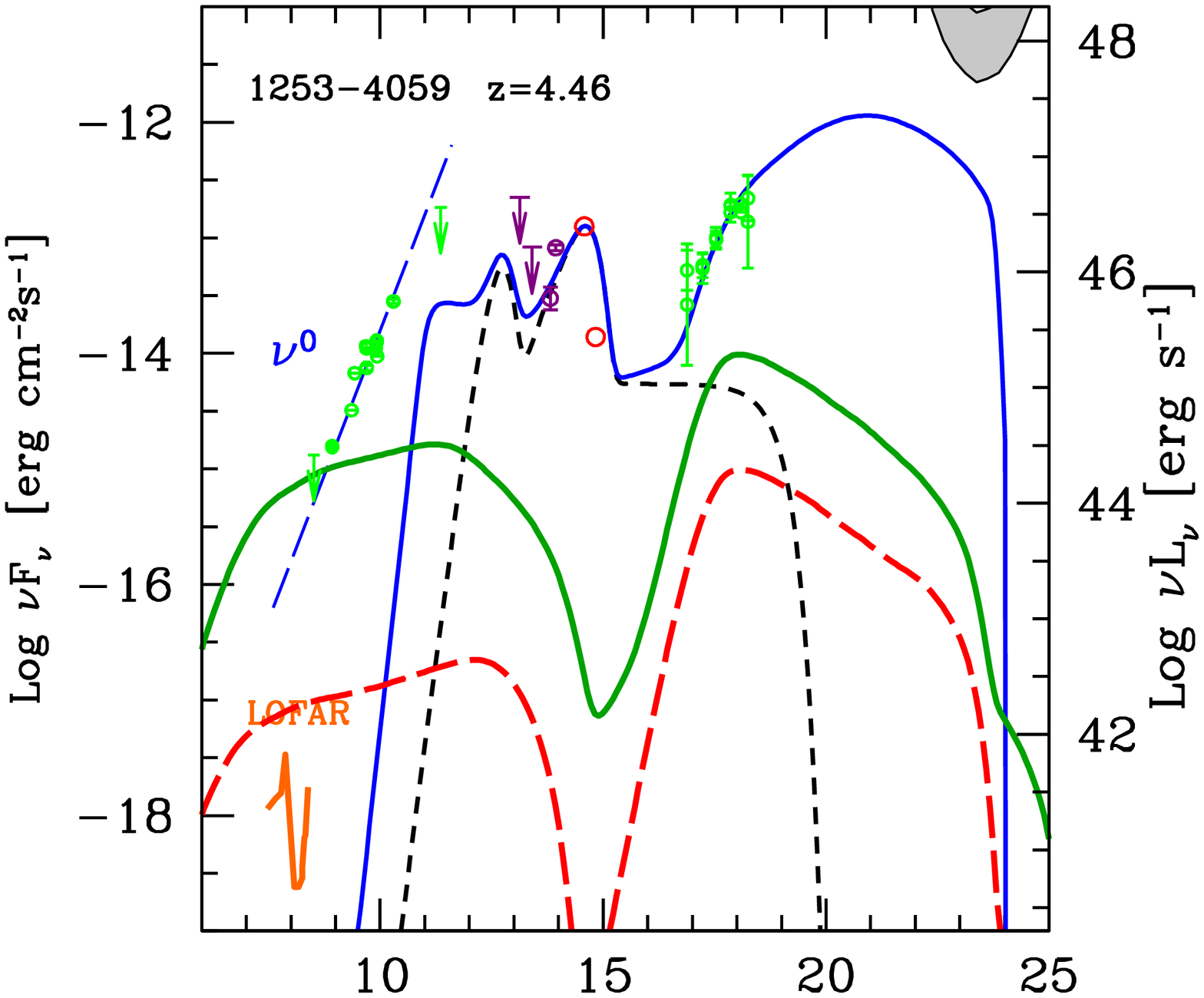,width=9cm,height=7.cm}
\vskip -1.3 cm
\psfig{file=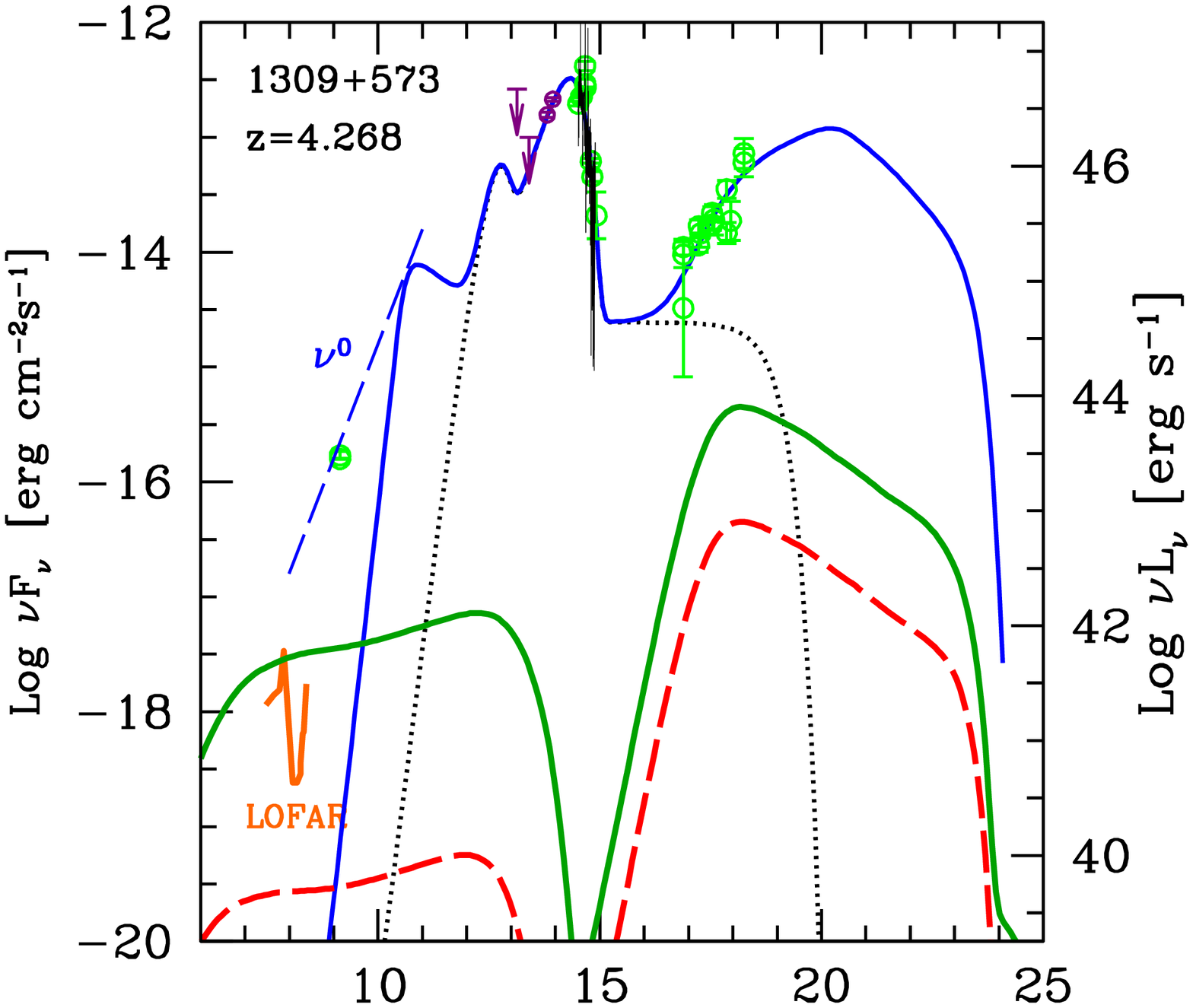,width=9cm,height=7.cm}
\vskip -1.3 cm
\psfig{file=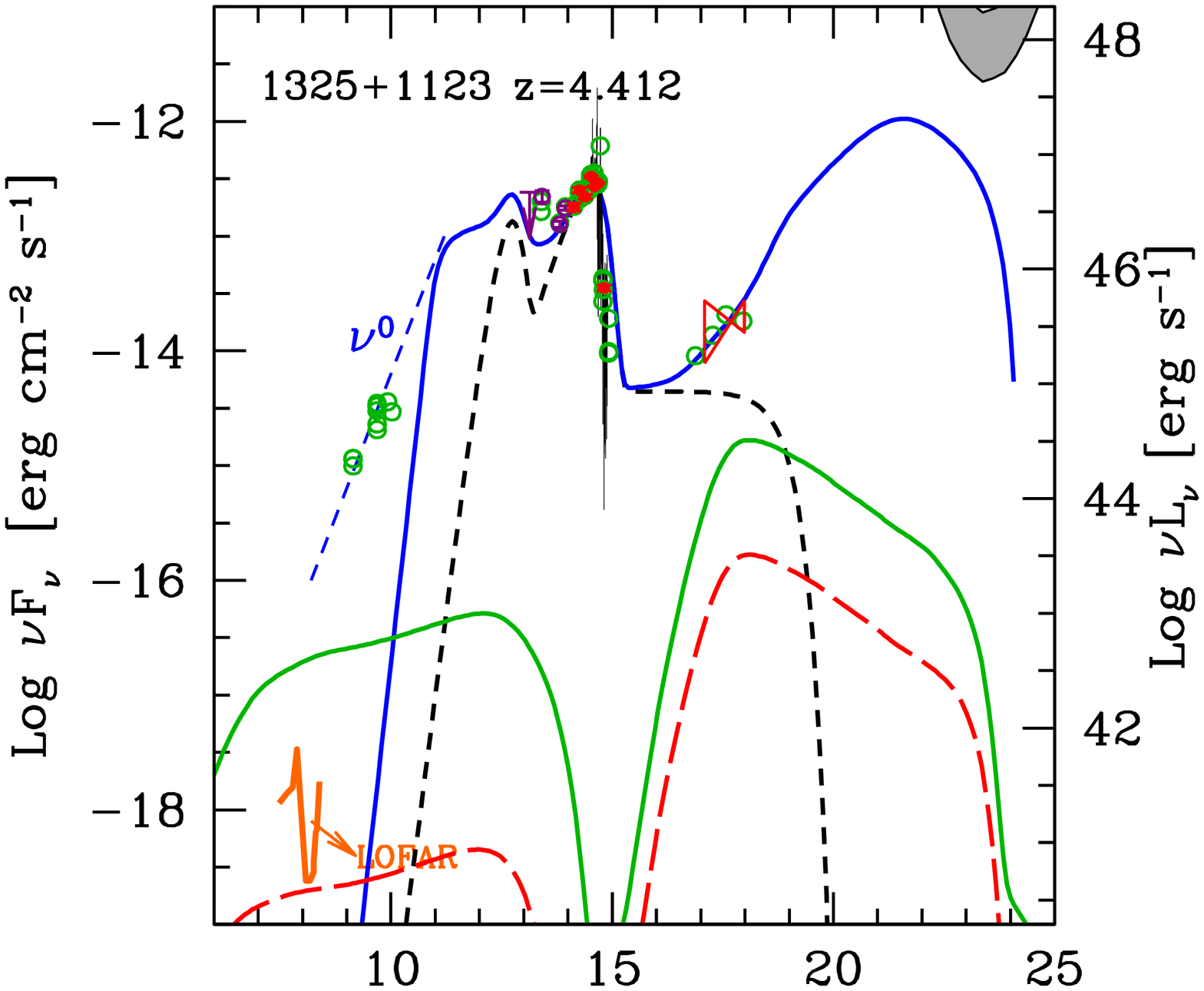,width=9cm,height=7.cm}
\vskip -1.3 cm
\psfig{file=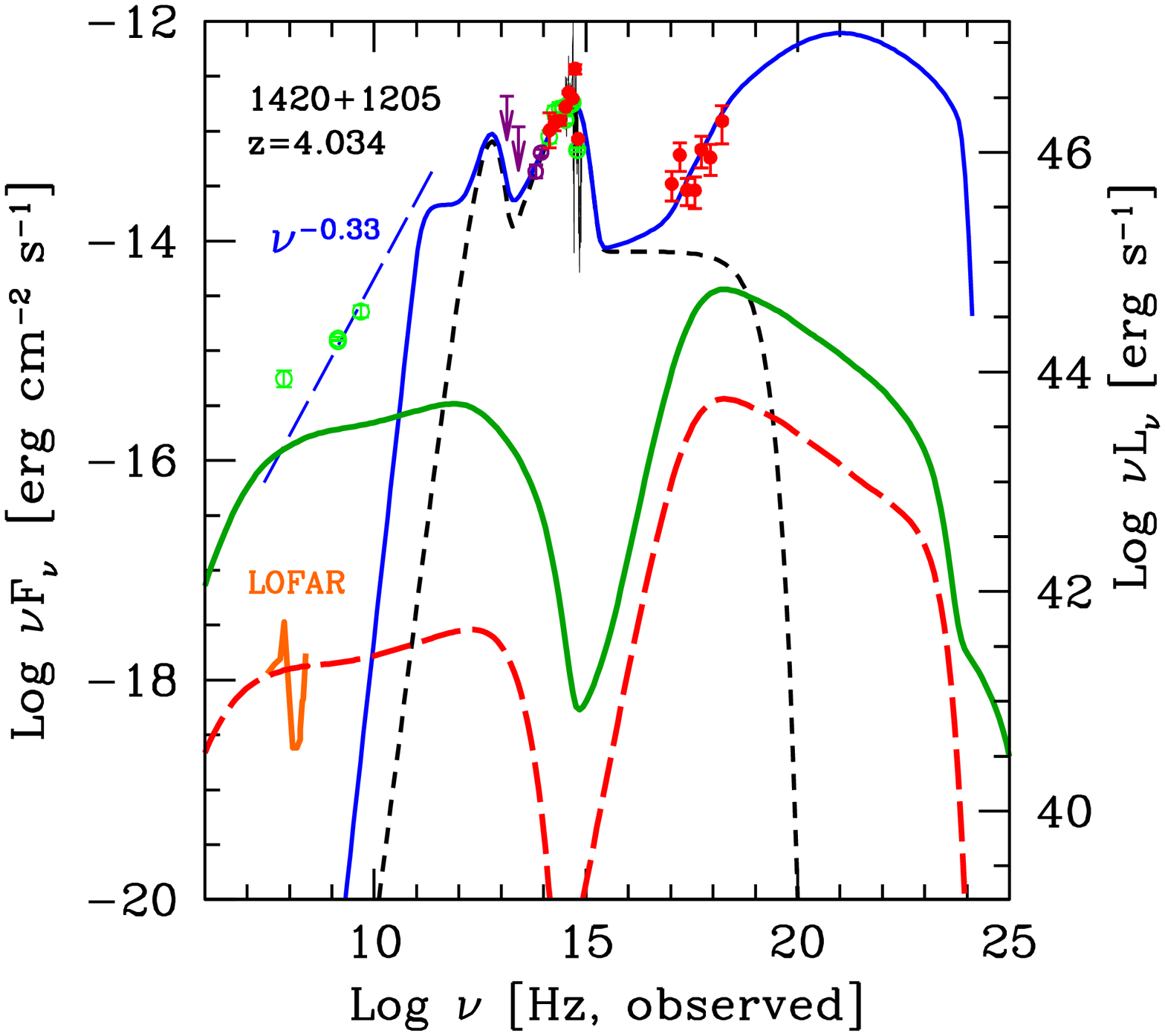,width=9cm,height=7.cm}
\vskip -0.7 cm
\caption{
{\it continue.}
SED of the $z>4$ blazars.
As in Fig. \ref{0324}, but showing only the sum of the hot spot and lobe components.
}
\label{sed}
\end{figure}

\setcounter{figure}{7}
\begin{figure}
\vskip -0.6 cm
\psfig{file=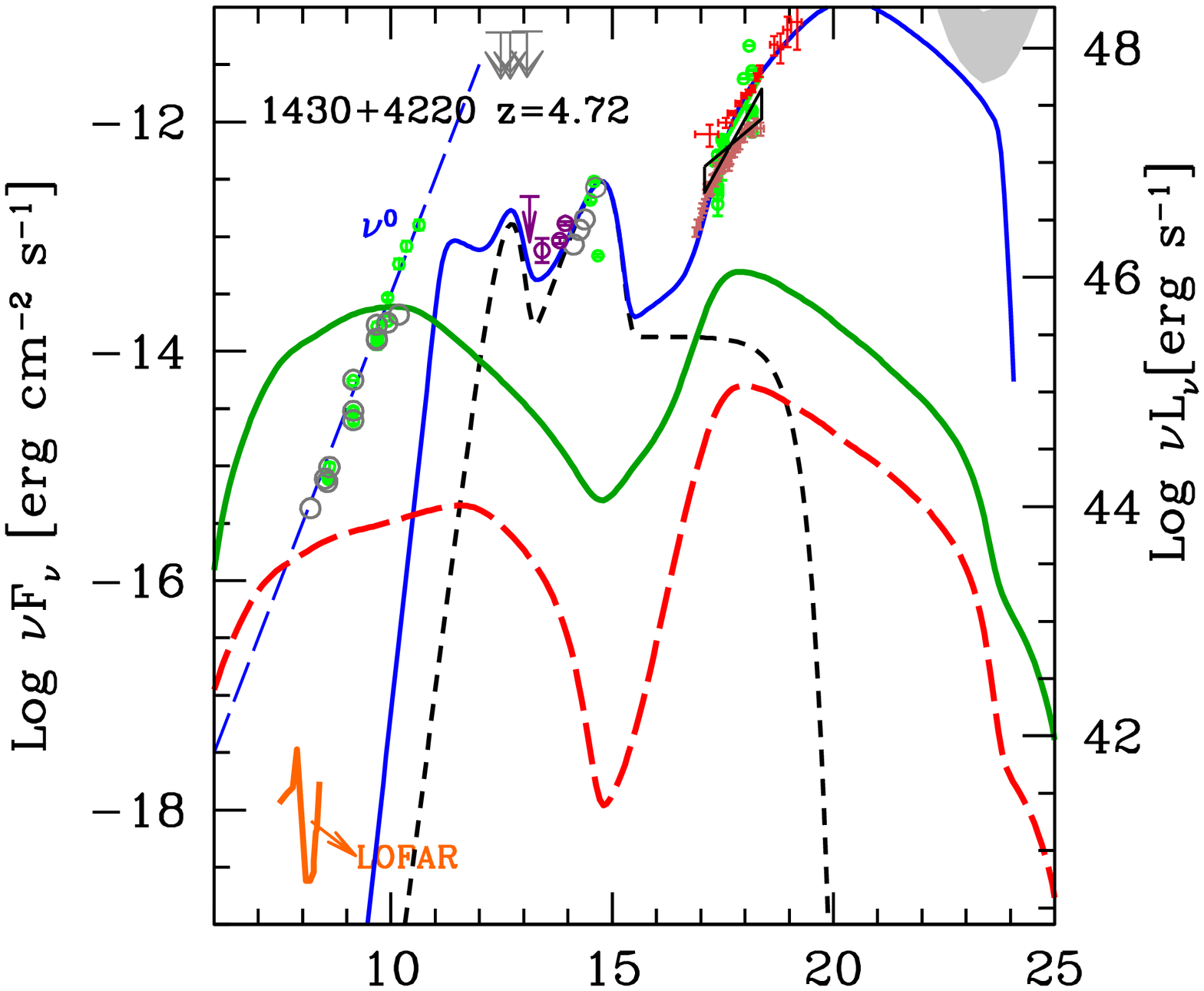,width=9cm,height=7.cm}
\vskip -1.3 cm
\psfig{file=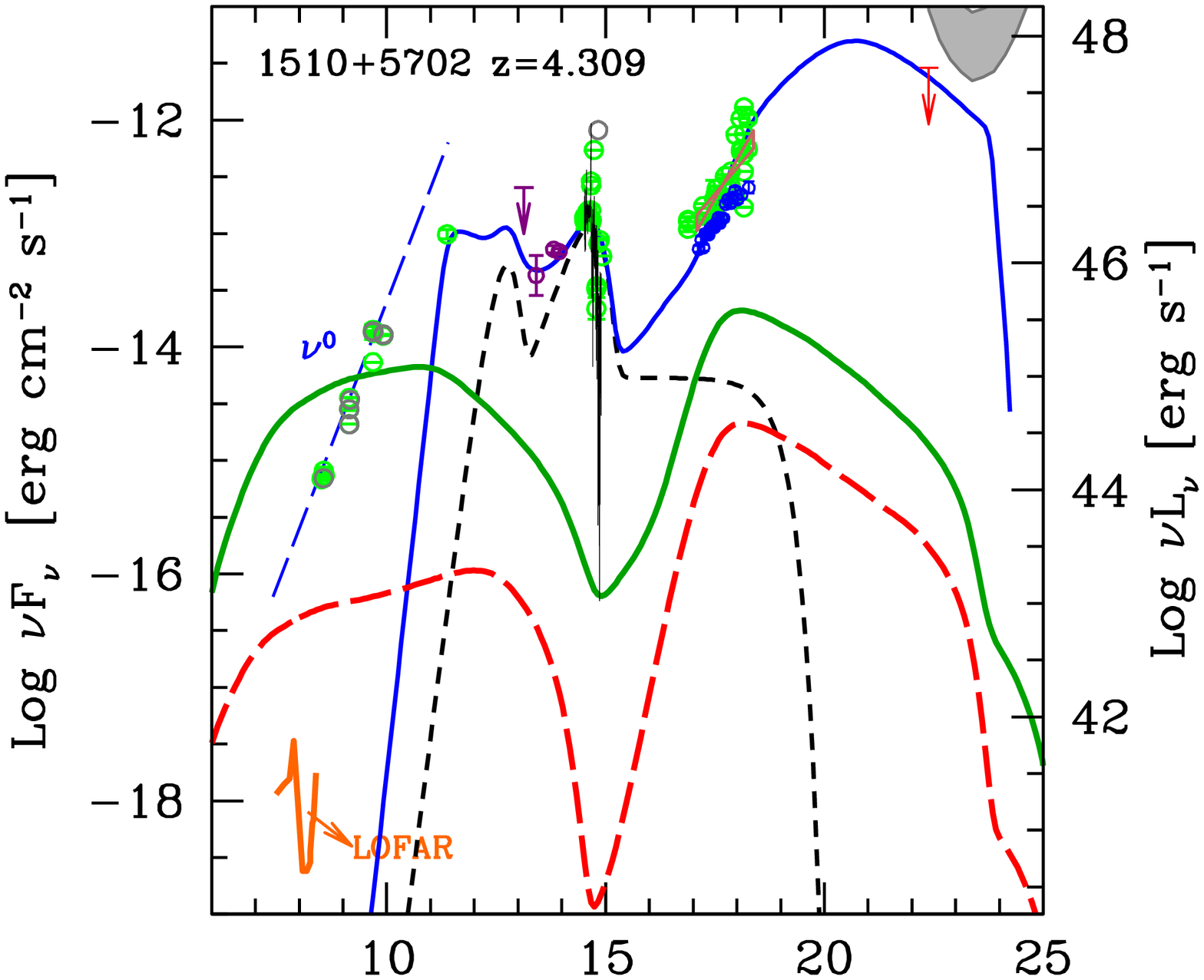,width=9cm,height=7.cm}
\vskip -1.3 cm
\psfig{file=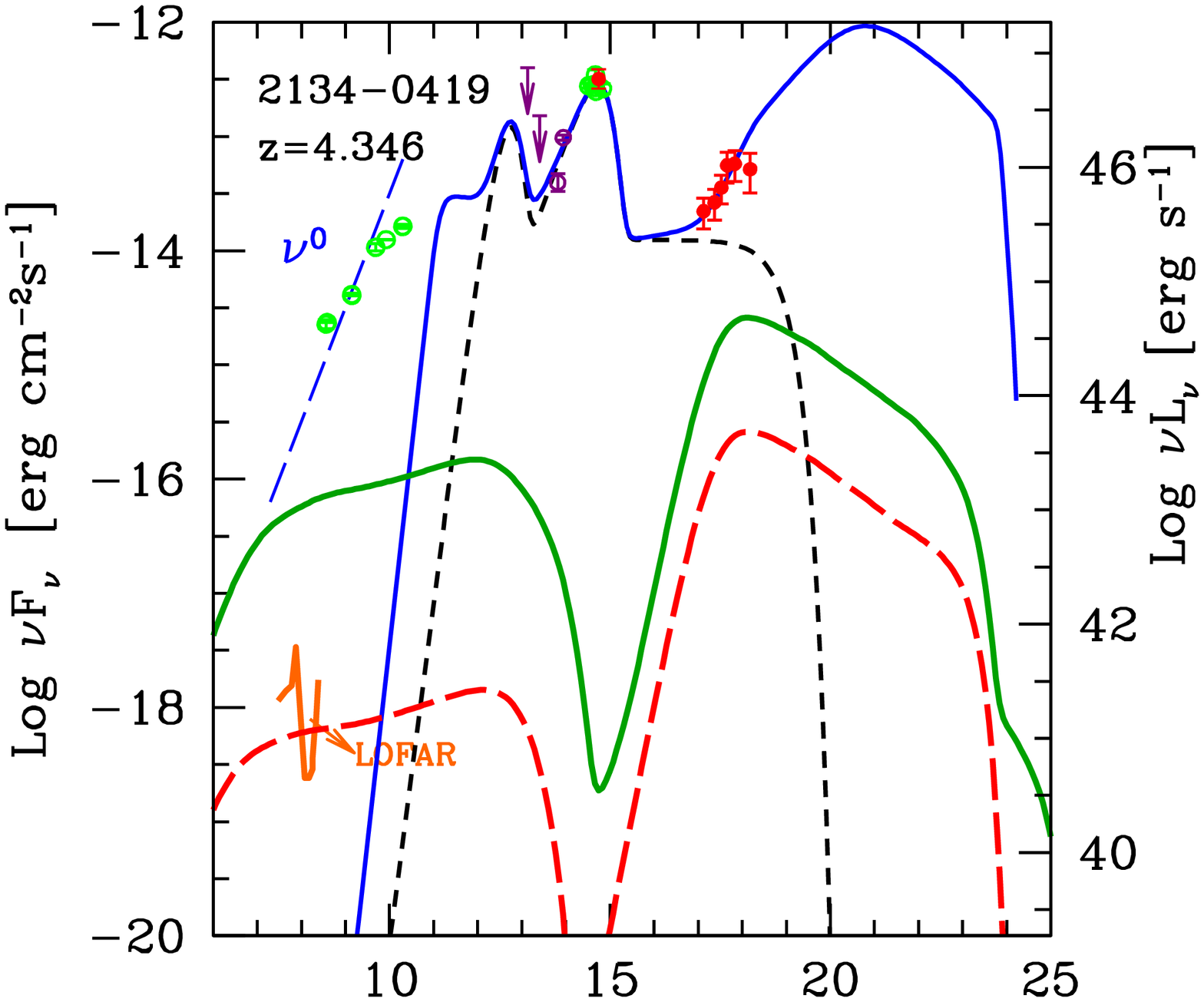,width=9cm,height=7.cm }
\vskip -1.3 cm
\psfig{file=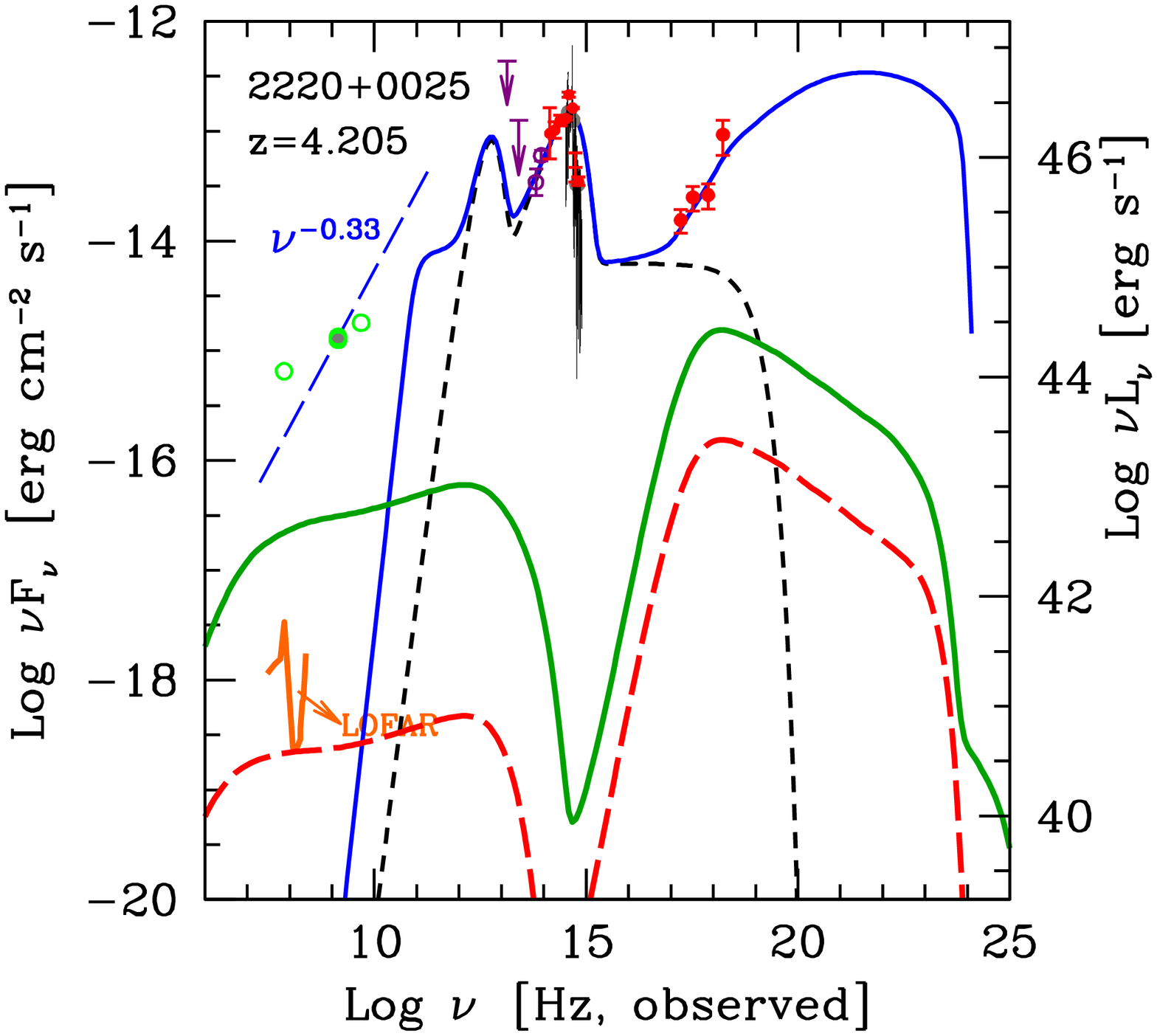,width=9cm,height=7.cm }
\vskip -0.7 cm
\caption{{\it continue.}
SED of the $z>4$ blazars.
As in Fig. \ref{0324}, but showing only the sum of the hot spot and lobe components.
}
\label{sed}
\end{figure}

We conclude that the available low--frequency radio data do not exclude the presence of 
extended emission in our blazars, and that  
CMB quenching is instrumental in reducing the radio lobe emission 
below current detection limits in all sky surveys.

Our  modelling relies on a number of assumptions on the fiducial values of key parameters, 
and the resulting SEDs are degenerate with respect to, e.g., the values of the 
magnetic field strength and of the source size. 
On the other hand, deeper observations by, e.g., LOFAR, should be able to detect 
the lobe radio emission in most sources, even in the conservative case $P_{\rm e,lobe}= 0.01 P_{\rm jet}$.
Should  the radio (and/or the X--ray) flux of the lobes be detected, the source 
physical parameters would be determined with much higher confidence. 
While in principle a simple low--frequency excess of the radio flux above the extrapolation of the flat 
jet spectrum marks the presence of a lobe, radio maps (and X--ray detections and, possibly, 
imaging; see \S 4.1 below) would allow us to measure the magnetic field and the injected power, 
and hence derive the total electron and magnetic energetics with minimum assumptions.

\section{Discussion and conclusions}

Previous works (e.g., Volonteri \etal 2011) found that the number density 
of radio--galaxies agrees with the expected parent population of blazars at $z\lsim 3$. 
At earlier epochs, however, there is a clear, strong deficit of radio--galaxies. 
One of the possible solutions for this puzzle is the $(1+z)^4$ enhancement of the CMB radiation 
energy density, which quenches the isotropic radio emission of the lobes while
boosts their X--ray luminosity.

We have then constructed a sample containing all known blazars  at $z>4$, modelling
their jet emission to ultimately assess the broad band emission of the lobes.
While current low frequency observations of high--$z$ blazars do not show any evidence of isotropically, 
optically thin synchrotron emission from extended lobes, our results indicate 
that such observations are simply not deep enough. 
We can predict that, for a range of plausible parameters, deeper pointings 
(by, e.g., LOFAR or by the expanded VLA) should be able to detect the lobe emission. 
At the same time, deep X--ray observations with high angular resolution (by, i.e., Chandra) have
the required sensitivity to detect and resolve the lobes in X--rays, as the estimated lobe angular size 
is large enough to avoid flux contamination from the point--like nuclear emission.

A measure of the lobe radio and X--ray fluxes, as well as of the 
lobe angular size, will allow us to estimate the physical parameters of the source with
no ambiguity and without invoking equipartition between the particles and the magnetic field.
Besides testing the validity of the CMB quenching scenario, this will give information about 
the energy in the proton component of the lobes, and
possibly the coherence length of the magnetic field.


According to our scenario, if the lobes were indeed large (namely, $R_{\rm lobe}>25$ kpc) and
in equipartition, then it would be difficult to detect them  
in any radio--galaxy in the radio band at high redshifts. 
The radiogalaxies at $z>4$ with detected 
extended radio emission apparently challenge our proposed CMB quenching scenario (see e.g.  
TN0924--2201 at $z=5.19$, van Breugel et al. 1999;  
6C 0140+326 at $z=4.41$, Rawlings et al. 1996;      
4C63.20 at $z=4.261$, Lacy et al. 1994a;            
PKS 1338--1942 at $z=4.11$, De Breuck et al. 1999). 
We defer a detailed study of these sources to a paper in preparation, yet we can anticipate 
that the reason of their strong radio emission most probably lies in the fact
that what we detect are the hot spots, not the lobes.
(see e.g. De Breuck et al. 2010 for an estimate of the lobe sizes). 	
The strong magnetic field of their hot spots
(of order of hundreds of $\mu$G), makes 
synchrotron losses competitive with the inverse Compton scattering off CMB photons.

Our results have a profound impact on the estimated fraction of radio loud high--$z$ AGNs. 
The jet emission in blazars is in fact easily visible even at high redshifts and is 
unaffected by the CMB quenching since: 
(i) jet emission is strongly enhanced by beaming, and 
(ii) the radio flux originates in compact regions where magnetic energy density 
dominates over the CMB. 
The beaming pattern of jet emission decreases fast for
increasing viewing angles: for $\Gamma=15$, the flux at $\theta_{\rm v}\sim 15^\circ$ 
is dimmed by $\sim$5 orders of magnitude compared to the one at $\theta_{\rm v}=3^\circ$.
This implies that misaligned, high--$z$ jets are invisible for the sensitivity
of current instruments.

\begin{figure}
\vskip -0.6 cm
\hskip -0.7 cm
\psfig{file=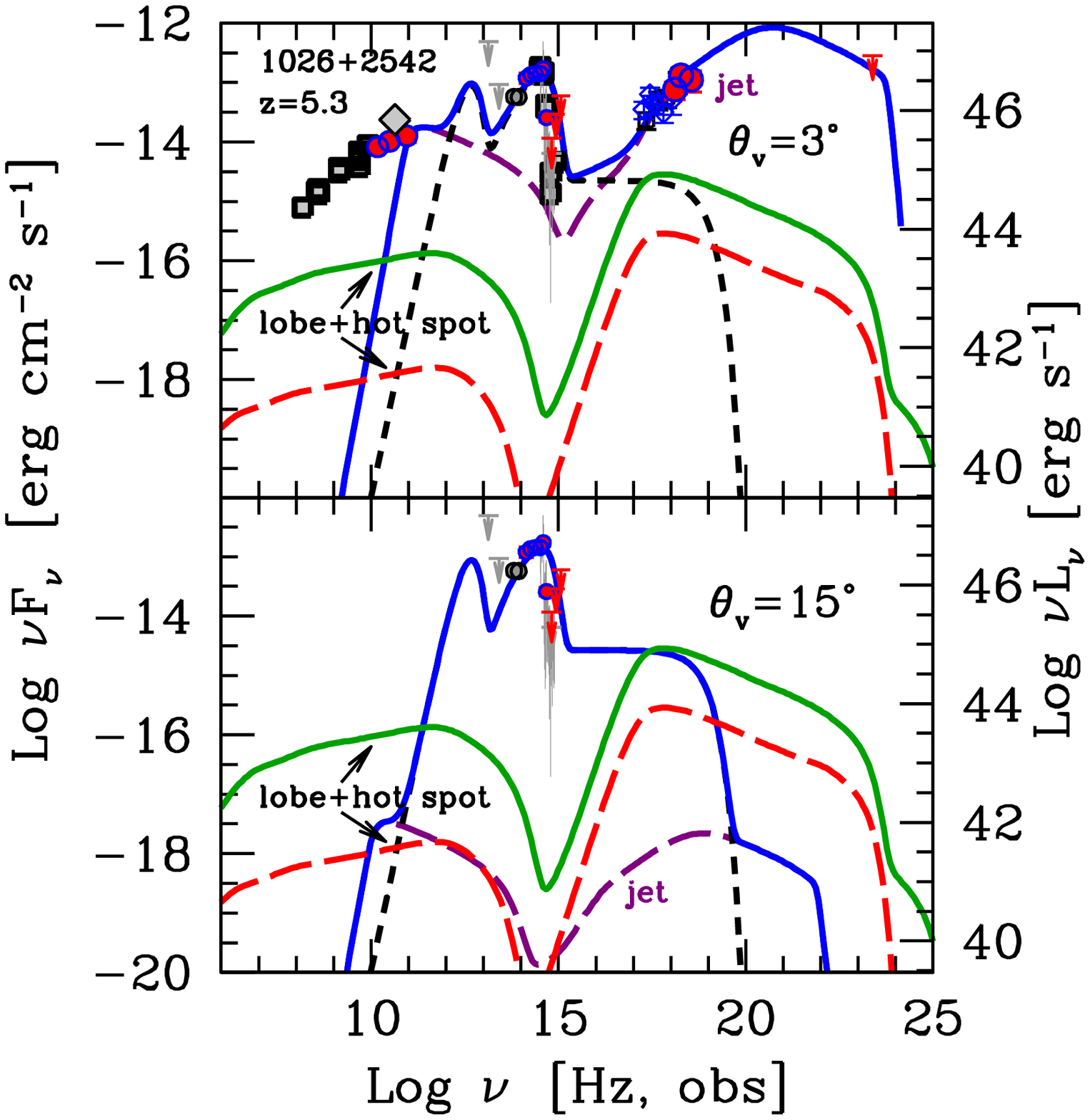,width=9.9cm,height=9.9cm }  
\vskip -0.5 cm
\caption{
{\it Top panel:}
the SED of the blazar 1026+2542 at $z=5.3$ is fitted with a model (parameters in
Tab. \ref{para}) with $\Gamma=15$ and $\theta_{\rm v}=3^\circ$, as in Fig. \ref{sed}.
Here we also show the contribution from the jet (long--dashed violet line), 
the thermal components
(disc, torus and corona, as a short--dashed black line) and their sum (solid blue line).
The solid green and the  dot--dashed red lines are the estimated lobe+hot spot emission
as in Fig. \ref{sed} (the two models differ from the assumed injected electron energy,
equal to 10\% of 1\% of $P_{\rm jet}$, respectively).
{\it Bottom panel:}
if the very same source is observed at $\theta_{\rm v}=15^\circ$ (i.e., only slightly off the beaming angle),
the jet emission (long--dashed red line) is so weak that it would be missed by all  
current surveys, while the isotropic components (disc, torus, corona and the lobes)
would remain the same.
{\it Current optical surveys, as the SDSS, would easily detect the source, and classify it
as a radio--quiet quasar.}
The lobe + hot spot emission, slightly below the sensitivity of FIRST (1 mJy at 1.4 GHz), 
could be detectable by LOFAR (see Fig. \ref{sed}); furthermore, pointed {\it Chandra} X--ray observations could
detect and resolve the X--ray emission of the lobes (note that 50 kpc at $z=5.3$ correspond to 
an angular size of 8.2 arcsec in the adopted cosmology).
These observations could establish the jetted nature of the source.
}
\label{1023}
\end{figure} 

Fig. \ref{1023} illustrates the point. 
We compare the SED of 1026+2542 ($z$=5.3) computed in \S 4 to the corresponding model as observed at 
viewing angle $\theta_{\rm v}=15^\circ$.
By modelling the hot spot+lobe emission as detailed in \S \ref{lobe} (green line in Fig. \ref{1023}), 
we find that the resulting isotropic radio flux would reach the mJy level at frequencies $\simeq $300 MHz, 
and would be much weaker at higher frequencies. 
As a result, a source like 1026+2542 whose jet is observed at $\theta_{\rm v}>15^\circ$ would be easily
detected in current optical survey, thanks to its quasi--isotropic accretion disc emission,
but would fail to enter radio--catalogs like FIRST, with its sensitivity limit of 1 mJy
at 1.4 GHz.
{\it Therefore, slightly misaligned high--$z$ jetted sources would be classified as radio--quiet}.
The very term ``radio--loud" becomes misleading. 
The idea that some high--$z$ radio--quiet AGNs could have extended X--ray emission,
revealing the presence of the jet, was first put forward by Schwartz (2002) 
analysing the X--ray data of three $z\sim$6 radio--quiet sources observed by {\it Chandra}. 
The short exposure of these observations seemed to suggest that the X--ray emission
was {\it extended}. 
However, the extended emission of one of them was then associated to a galaxy (Ivanov 2004),
and this was confirmed by longer Chandra observations by Schwartz \& Virani (2004).
Brandt et al. (2002) did not confirm the extended nature of the X--ray flux observed in the other two sources.

Fig. \ref{1023} also shows that most of the lobe radiative losses are instead in the X--ray band.
Electrons emitting above $\simeq $1 keV are cooling in less than one source crossing time,
and therefore are responsible for releasing almost the entire injected power 
$P_{\rm e}^{\rm lobe}\sim 0.1 P_{\rm jet}=5\times 10^{45}$ erg s$^{-1}$. 
Lobes are then radio--quiet and X--ray loud. 
Interestingly, in the 2--10 keV X--ray band 
the extended lobe and the nuclear accretion disk corona emissions can be comparable, as shown in Fig. \ref{1023}. 
As a linear proper dimension of 100 kpc corresponds at $z=5$ to an angular size of 16 arcsec, sub--arcmin X--ray 
imaging is required to disentangle the two components.

\section*{Acknowledgments}
We thank the referee, John Wardle, for very useful comments and
criticisms, that helped to improve the paper.
This research made use of the NASA/IPAC Extragalactic Database (NED) which is operated 
by the Jet Propulsion Laboratory, Caltech, under contract with NASA.
Part of this work is based on archival data, software or online 
services provided by the ASI Science Data Center (ASDC).


\begin{thebibliography}{}


\bibitem[]{} Ackermann M., Ajello M., Allafort A. et al., 2011, ApJ, 743, 171 

\bibitem[]{} Bassett L.C., Brandt W.N., Schneider D.P., Vignali C., Chartas G., Garmire G.P., 2004, AJ, 128, 523

\bibitem[]{} Becker R.H., White R.L. \& Helfand D.J., 1995, ApJ, 450, 559

\bibitem[]{} Belsole E., Worrall D.M., Hardcastle M.J., Birkinshaw M. \& Lawrence C.R.
             2004, MNRAS, 352, 924 


\bibitem[]{} Brandt W.N., Schneider D. P., Fan X. et al., 2002, ApJ, 569, L5

\bibitem[]{} Carilli C.L. \& Taylor G.B., 2002, ARA\&A, 40, 319

\bibitem[]{} Carilli C.L., R\"ottgering H.J.A., van Ojik R., Miley G.K. \& 
             van Breugel W.J.M., 1997, ApJSS, 109, 1 
         
\bibitem[]{} Celotti A. \& Fabian A.C., 2004, MNRAS, 353, 523

\bibitem[]{} Croston J.H., Birkinshaw M., Hardcastle M.J. \& Worrall D.M., 2004, MNRAS, 353, 879

\bibitem[]{} Croston J.H., Hardcastle M.J., Harris D.E., Belsole E., Birkinshaw M. \& Worrall D.M., 2005, ApJ, 626, 733

\bibitem[]{} De Breuck C., van Breugel W., Minniti D., Miley G., R\"ottgering H., 
             Stanford S.A. \& Carilli C., 1999, A\&A, 352, L51 

\bibitem[]{} De Breuck C., Seymour N., Stern D. et al., 2010, ApJ, 725, 36 

\bibitem[]{} de Vries W.H., Becker R.H. \& White R.L., 2006, AJ, 131, 666 

\bibitem[]{} Fabian A.C., Celotti A., Iwasawa K., Ghisellini G., 2001a, MNRAS, 324, 628

\bibitem[]{} Fabian A.C., Celotti A., Iwasawa K., McMahon R.G., Carilli C.L., Brandt W.N., 
             Ghisellini G., Hook I. M., 2001b, MNRAS, 323, 373

\bibitem[]{} Fabian A.C., Walker S.A., Celotti A., Ghisellini G., Mocz P., 
             Blundell K.M. \& McMahon R.G. 2014, MNRAS, 442, L81


\bibitem[]{} Ghisellini G., 1999, AN, 320, 232 

\bibitem[]{} Ghisellini G. \& Tavecchio F., 2009, MNRAS, 397, 985  

\bibitem[]{} Ghisellini G. \& Tavecchio F., 2010, MNRAS,  409, L79  

\bibitem[]{} Ghisellini G., Celotti A., Tavecchio F., Haardt F. \& Sbarrato T.,
             2014a, MNRAS, 438, 2694  

\bibitem[]{} Ghisellini G., Sbarrato T., Tagliaferri G., Foschini L., Tavecchio F., Ghirlanda G., 
             Braito V. \& Gehrels N., 2014b, MNRAS, 440, L111 

\bibitem[]{} Ghisellini G., Tavecchio F., Maraschi L., Celotti A. \& Sbarrato T., 2014c, Nature, 515, 376

\bibitem[]{} Ghisellini G., Tagliaferri G., Sbarrato T. \&  Gehrels N., 2015, MNRAS, 450, L34 

\bibitem[]{} Ghisellini G. \& Tavecchio F., 2015, MNRAS, 448, 1060

\bibitem[]{} Gobeille D.B., Wardle J.F.C. \& Cheung C.C., 2014, preprint (arXiv: 1406.4797)  

\bibitem[]{} Haiman Z., Quataert E. \& Bower G.C., 2004, ApJ, 612, 698 

\bibitem[]{} Hardcastle M.J. \& Krause M.G.H., 2014, MNRAS, 443, 1482

\bibitem[]{} Hook I.M. \& McMahon R.G., 1998,  MNRAS, 294, L7

\bibitem[]{} Hook I.M., McMahon R.G., Shaver P.A. \& Snellen I.A.G., 2002, A\&A, 391, 509


	


\bibitem[]{} Ivanov V.D., 2002, A\&A, 389, L37

\bibitem[]{} Kratzer R. M.\& Richards G.T., 2015, AJ, 149, 61

\bibitem[]{} Lacy M., Miley G., Rawlings S. et al., 1994, MNRAS, 271, 504

\bibitem[]{} Massaro F., Giroletti M., Paggi A., D'Abrusco R., Tosti G. \& Funk S., 2013, ApJS, 208, 15

\bibitem[]{} Mc Greer I.D., Helfand  D.J. \& White R.L., 2009, AJ, 138, 1925

\bibitem[]{} Mocz P., Fabian A.C. \& Blundell K.M., 2011, MNRAS, 413, 1107

\bibitem[]{} Nemmen R.S., Georganopoulos M., Guiriec S., Meyer E.T., Gehrels N. \& Sambruna R.M.,
            2012, Science, 338, 1445


\bibitem[]{} Rawlings S., Lacy M., Blundell K.M., Eales  S.A., Bunker A.J. 
            \& Garrington S.T., 1996, Nature, 383, 502 
             
\bibitem[]{} Rees M.J., 1978, MNRAS, 184, P61
 
\bibitem[]{} Romani R.W., Sowards--Emmerd D., Greenhill L., Michelson P., 2004, ApJL, 610, L9

\bibitem[]{} Sbarrato T., Ghisellini G., Nardini M., et al., 2012, MNRAS, 426, L91 

\bibitem[]{} Sbarrato T., Ghisellini G., Nardini M., Tagliaferri G., Greiner J., Rau A. \& Schady P.,
             2013a, MNRAS, 433, 2182 
             
\bibitem[]{} Sbarrato T., Tagliaferri G., Ghisellini G. et al., 2013b, ApJ, 777, 147 

\bibitem[]{} Sbarrato T., Ghisellini G., Tagliaferri G., Foschini L., Nardini M., Tavecchio F., 
             Gehrels N., 2015, MNRAS, 446, 2483 

\bibitem[]{} Shakura N.I. \& Sunyaev R.A., 1973 A\&A, 24, 337    %

\bibitem[]{} Shaw M.S., Romani R.W., Cotter G. et al.,  2012, ApJ, 748, 49 

\bibitem[]{} Shen Y., Richards G.T., Strauss M.A. et al., 2011, ApJS, 194, 45  
	

\bibitem[]{} Schoenmakers A.P., Mack K.-H., de Bruyn A.G., R\"ottgering H.J.A., Klein U. \& van der Laan H.,
             2000, A\&AS, 146, 293 
	
\bibitem[]{} Schwartz D.A., 2002, ApJ, 571, L71

\bibitem[]{} Schwartz D.A. \& Virani S.N., 2004, ApJ, 615, L21


\bibitem[]{} Spada M., Ghisellini G., Lazzati D. \& Celotti A., 2001, MNRAS, 325, 1559)

\bibitem[]{} van Breugel W., De Breuck C. Stanford S. A., Stern D., R\"ottgering H. \&  
             Miley G. 1999, ApJ, 518, L61
              
\bibitem[]{} van Haarlem, Wise M.W., Gunst A.W. et al., 2013, A\&A, 556, 2  

\bibitem[]{} Van Velzen S., Falcke H. \& K\"ording E., 2015, MNRAS, 446, 2985

\bibitem[]{} Volonteri M., Haardt F., Ghisellini G. \& Della Ceca R., 2011, MNRAS, 416, 216

\bibitem[]{} Yuan W., Matsuoka M., Wang T., Ueno S., Kubo H., Mihara T., 2000, ApJ, 545, 625

\bibitem[]{} Yuan W., Fabian A.C., Worsley M.A., McMahon R.G., 2006, MNRAS, 368, 985

\bibitem[]{} Wilson A.S., Yoing A.J \& Shopbell P.L., 2000, ApJ, 544, L27 

\bibitem[]{} Wu J., Brandt W.N., Miller B.P., Garmire G.P., Schneider D.P., Vignali C., 2013, ApJ, 763, 109 


\end{thebibliography}
 \end{document}